\documentclass[%
 aps,prl,twocolumn,longbibliography,reprint,superscriptaddress,
 amsmath,amssymb,
floatfix,
]{revtex4-2}

\usepackage{blindtext}
\usepackage{graphicx}% Include figure files
\usepackage{dcolumn}% Align table columns on decimal point
\usepackage{bm}% bold math
\usepackage{marginnote}
\usepackage{inputenc}

\usepackage{url}

\usepackage{hyperref,color,pgfplots,siunitx,multirow,braket,adjustbox,float}
\usetikzlibrary{plotmarks}
\pgfplotsset{compat=1.13} %ShareLatex compatibility error fix

\hypersetup{
    unicode=true, % non-Latin characters in Acrobat?s bookmarks
    plainpages=false,
    colorlinks=true,% false: boxed links; true: colored links
    linkcolor=blue,% color of internal links
    citecolor=blue,% color of links to bibliography
    filecolor=blue,% color of file links
    urlcolor=blue% color of external links
}

\usepackage{txfonts} % Nicer fonts?

\usepackage{xargs}
\usepackage[colorinlistoftodos,prependcaption,textsize=footnotesize]{todonotes}
 \newcommandx{\unsure}[2][1=]{\todo[linecolor=red,backgroundcolor=red!25,bordercolor=red,#1]{#2}\ }
 \newcommandx{\change}[2][1=]{\todo[linecolor=blue,backgroundcolor=blue!25,bordercolor=blue,#1]{#2}\ }
 \newcommandx{\info}[2][1=]{\todo[linecolor=yellow,backgroundcolor=yellow!25,bordercolor=yellow,#1]{#2}\ }
 \newcommandx{\add}[2][1=]{\todo[linecolor=green,backgroundcolor=green!25,bordercolor=green,#1]{#2}\ }

\usepackage[normalem]{ulem}

\usepackage{upgreek}

\usepackage{calc}

\begin{document}

\title{Phonon Signatures in Photon Correlations}

\author{Ben S. Humphries}
\affiliation{School of Chemistry, University of East Anglia, Norwich Research Park, Norwich, NR4 7TJ, United Kingdom}
\author{Dale Green}
\affiliation{School of Chemistry, University of East Anglia, Norwich Research Park, Norwich, NR4 7TJ, United Kingdom}
\author{Magnus O. Borgh}
\email{M.Borgh@uea.ac.uk}
\affiliation{Physics, Faculty of Science, University of East Anglia, Norwich NR4 7TJ, United Kingdom}
\author{Garth A. Jones}
\email{garth.jones@uea.ac.uk}
\affiliation{School of Chemistry, University of East Anglia, Norwich Research Park, Norwich, NR4 7TJ, United Kingdom}

\begin{abstract}
We show that the second-order, two-time correlation functions for phonons and photons emitted from a vibronic molecule in a thermal bath result in bunching and anti-bunching (a purely quantum effect), respectively.  Signatures relating to phonon exchange with the environment are revealed in photon-photon correlations. We demonstrate that cross-correlation functions have a strong dependence on the order of detection giving insight into how phonon dynamics influences the emission of light. This work offers new opportunities to investigate quantum effects in condensed-phase molecular systems. 
    % Up to 600 characters (495/600), Last edited Jan 19. Now 519 without spaces, 596 with (14:00 Jan 19)
\end{abstract}

\maketitle
% Fig1:331 Fig2:209 Fig3:225 Body: 2600 (with captions)
% Fig1:272 Fig2:192 Fig3:208 Body: 2600 (with captions)
%Eqn no: 7, Eqn words = 7*16 = 112
% Up to 3750 words (3477/3750) max, min (3384/3750) Updated Jan 31

\begin{figure*}[ht]
\centerline{\includegraphics[width=\textwidth]{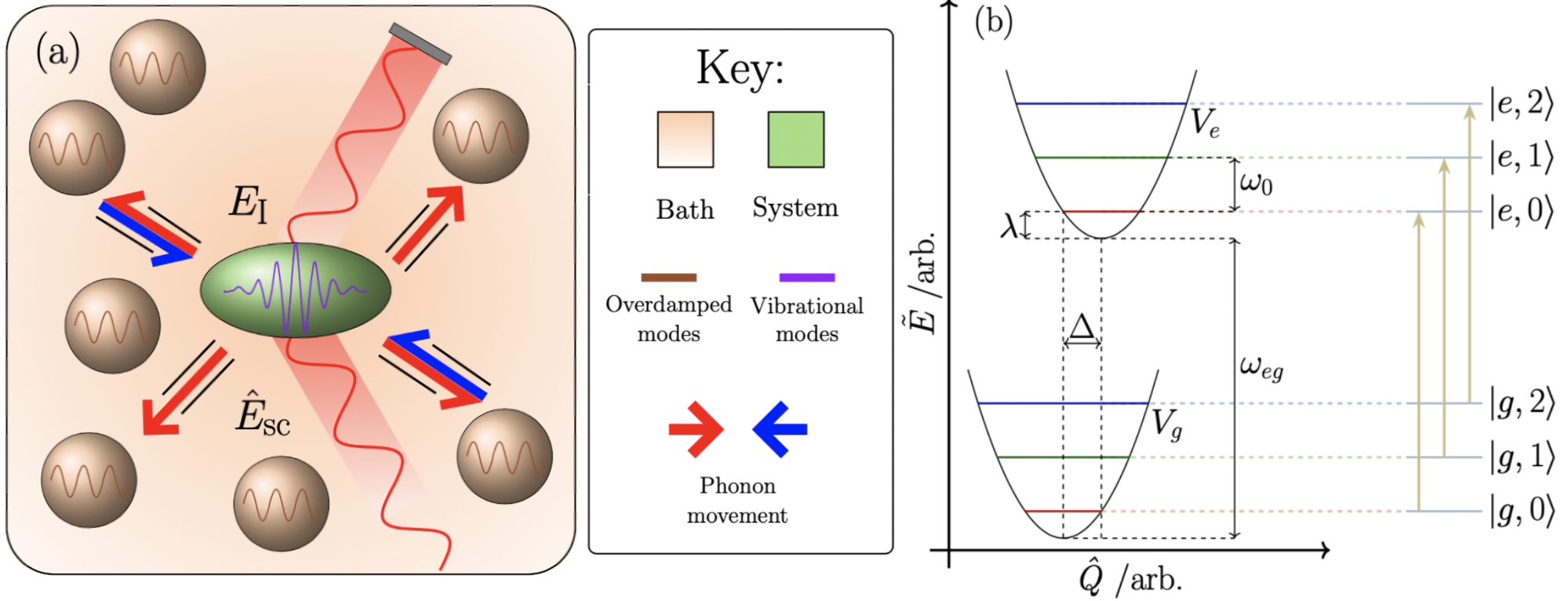}}

\caption{(a) Schematic of the molecule coupled to bath modes and driven by laser field $E_{I}$, resulting in the scattered field $\hat{E}_{\mathrm{sc}}$. Phonon movement between system and environment indicated by arrows. (b) Diabatic energy levels, with excited state displacement $\Delta$, system reorganization energy $\lambda$, fundamental transition frequency $\omega_{eg}$, and system mode frequency $\omega_0$. Corresponding adiabatic levels on the far right.} \label{schematic}
\end{figure*}

%\blue{\textbf{Height: \the\graphicheight}}
%\blue{\textbf{Width: \the\graphicwidth}}
% this is [300/(0.5*AR)]+40 for double column
% h: 196.58628pt w: 510.0pt AR = 2.59428074...
% Fig3 words: 272 + 59 caption 331

%\pagebreak
Correlation measurements of photon emission provide powerful tools for demonstrating quantum effects. Among the most striking examples is the experimental discovery of anti-bunching in the photon emission of fluorescing atoms~\cite{Kimble1977,Kimble1978}, which provided the first direct demonstration of the quantum properties of light~\cite{Walls1979}. Anti-bunching~\cite{Paul1982} is the phenomenon whereby the emission of a second photon immediately after a first is suppressed. The joint probability of detecting two photons at time $t$ and $t+\tau$ is quantified by the second-order correlation function~\cite{Carmichael1976}. While in classical emission, this function may have a maximum for $\tau=0$, second-order correlations falling off as $\tau\to0$ can appear only as a purely quantum phenomenon. The second-order correlation function thus forms a powerful statistical tool that has been used to study fundamental properties of photon emission and photon-mediated interactions, for example bunching and anti-bunching in transmission through waveguides~\cite{Zheng2012} and in emission from plasmonic nanojunctions~\cite{Avriller2021},  pattern formation in photoinduced nucleation~\cite{Ishida2012}, photon-blockade effects~\cite{Imamoglu1997} in optical cavities~\cite{Birnbaum2005,Lang2011,Deng2017,Gu2021,Guo2022} (including modified response at strong coupling~\cite{Ridolfo2012}), atomic arrays~\cite{Cidrim2020, Williamson2020a}, as well as superatom behavior in ensembles of quantum emitters~\cite{Williamson2020a,Chen2022}. Higher-order correlations can further reveal two-photon blockade~\cite{Feng2021}.

In all of these examples, the correlations are exhibited by emitted light. However, the usefulness of quantum correlation functions extends beyond the study of photons. Similar techniques have also been applied to describe phonon blockade in opto-mechanical~\cite{HassaniNadiki2022,Li2022} and spin-mechanical~\cite{Wang2022,Yin2022} systems.
Intriguingly, second-order cross-correlation functions between non-identical particles and quasiparticles can reveal, for example, photon-magnon blockade in a ferrimagnetic material coupled to a microcavity~\cite{Moslehi2022},
and photon-phonon bunching and anti-bunching in a qubit-phonon-plasmon system under strong coupling~\cite{Ma2022}.

In recent years, there has been significant interest in the nature of electronic and vibrational coherence in condensed-phase resonance energy transfer in molecular systems~\cite{Scholes2017,Jones2019}. However, its exact nature, whether classical or quantum, remains controversial~\cite{Miller2012,Mancal2020}. This is because experiments typically used to
investigate this measure optical responses
resulting from an induced macroscopic polarization, which is
a classical property~\cite{Mukamel1995}. Quantum-optical techniques, such as correlation measurements~\cite{Kurt2021,Ban2019},  offer an avenue to investigate genuine quantum effects in molecular systems directly.

Here we develop an open quantum system model for a vibronic molecule driven by a continuous monochromatic laser field (Fig.~\ref{schematic}). We include both vibrational and electronic degrees of freedom as well as coupling to a thermal environment. We calculate photon-photon, phonon-phonon and photon-phonon correlations during cyclic pumping while the molecule undergoes vibrational relaxation (VR). The key result is that signatures relating to phonon exchange with the environment, which are revealed in phonon-phonon correlations, can be accessible through the measurement of photon emission. By examining photon-phonon cross-correlation functions~\cite{Moslehi2022,Ma2022,Abo2022,Ban2019,Kalaga2022}, we explain how phonon dynamics influences the emission of light. Measurements of such features could help elucidate the impact of vibrational excitations on the quantum nature of light-matter interaction processes in systems ranging from subwavelength molecular arrays~\cite{Holzinger2022} to large organic molecules~\cite{Pruchyathamkorn2020}.

We consider a model system consisting of a simple molecule with two electronic levels, each with a set of $N$ vibrational states, coupled to an infinite ensemble of overdamped quantum harmonic oscillator modes~\cite{Breuer2015,Breuer2002,Weiss1993}. Similar models have previously been used to investigate the role of a vibrational environment on the open quantum system dynamics of molecules~\cite{Green2019,Humphries2022}.

The total Hamiltonian is the formal sum
\begin{equation}
    \label{H_tot}    \hat{H}_{\mathrm{tot}}=\hat{H}_{S}+\hat{H}_{B}+\hat{H}_{SB}+\hat{H}_{SF},
\end{equation}
which describes the system, the bath, and the system-bath interaction, as well as the coupling between the system and the photon field.
The molecule is driven by a continuous, monochromatic laser field $\boldsymbol{E}_{I}=E_0\boldsymbol{\mathbf{e}} \ \large{[}\exp(i\omega_{eg} t/\hbar)+\exp(-i\omega_{eg} t/\hbar)\large{]}$, with polarization $\mathbf{e}$ and frequency $\omega_{eg}$ resonant with the fundamental transition between the ground ($\ket{g}$) and excited ($\ket{e}$) electronic levels,
such that
$\hat{H}_{SF}=-\boldsymbol{\hat{\mu}}\cdot\boldsymbol{E}_{I}(r,t)$.
Here $\boldsymbol{\hat{\mu}}=\mu_{eg}\mathbf{d}(\ket{e}\!\bra{g}+\ket{g}\!\bra{e})$ is the transition dipole moment of magnitude $\mu_{eg}$ and direction $\mathbf{d}$, such that $\hat{H}_{SF}=E_0\mu_{eg}\mathbf{e}\cdot\mathbf{d}\large{[}\exp(i\omega_{eg} t/\hbar)+\exp(-i\omega_{eg} t/\hbar)\large{]}\large{[}\ket{e}\!\bra{g}+\ket{g}\!\bra{e}\large{]}$~\cite{Mollow1969}. 
Driving by the laser results in stimulated photon emission yielding a scattered electric field with positive-frequency component 
$\hat{E}_{\mathrm{sc}}^{+}\sim \exp(i\omega_{eg}t/\hbar)\ket{e}\!\bra{g}$~\cite{Glauber1963}.

The vibrational degrees of freedom are described by 
    \begin{equation}
        \hat{H}_{S}= |g\rangle \hat{h}_{g} \langle g|+|e\rangle \hat{h}_{e}\langle e|. \label{MukHs}
    \end{equation}
The nuclear Hamiltonians for the ground and excited electronic states are, respectively,  defined by 
 \begin{eqnarray} 
        \hat{h}_{g}&=& \hbar\omega_0\left(\hat{b}^{\dagger} \hat{b}+\tfrac{1}{2}\right), \label{2ndHg} \\
        \hat{h}_{e}&=&  \hbar(\omega_{eg}+\lambda)+\hbar\omega_0\left(\hat{b}^{\dagger} \hat{b}-\sqrt{\frac{\lambda}{\omega_0}}(\hat{b}+\hat{b}^{\dagger})+\tfrac{1}{2}\right),\label{2ndHe}
    \end{eqnarray}
where $\omega_0$ is the system mode frequency, $\hat{b}^{\dagger}$ and $\hat{b}$ are system phonon creation and annihilation operators, corresponding to the vibrational states of the molecule, and $\lambda$ is the system reorganization energy~\cite{Egorova2007}. This model is constructed in a diabatic (D) basis that separates the vibrational levels from the electronic states, leading to explicit off-diagonal couplings. As a result, the electronic excited state, Eq.~\eqref{2ndHe}, appears displaced by $\Delta=\sqrt{2\lambda\omega_0^{-1}}$  relative to the electronic ground state [see also Fig.~\ref{schematic}(b)]. This displacement accounts for the change in the equilibrium geometry of the electronic excited state. Note that an increase in the system reorganization energy, $\lambda$, corresponds to an increased displacement, $\Delta$. The excited potential also experiences an energy shift $\hbar\omega_{eg}$ corresponding to the fundamental transition.

Instead of working in the diabatic basis, used above for conceptual clarity in the 
construction of the Hamiltonian, one can obtain adiabatic (A) eigenstates,  $\ket{g,0},\ket{g,1},\ldots,\ket{e,0},\ket{e,1},\ldots$ (see Fig.~\ref{schematic}), by diagonalising $\hat{H}_{\mathrm{tot}}$ through a unitary transformation $\hat{H}_{\mathrm{tot}}^{A}=(\hat{U}^{AD})^{\dagger}\hat{H}^{D}_{\mathrm{tot}}\hat{U}^{AD}$~\cite{Humphries2022,supplement}. 
The presence of the system reorganization energy in Eq.~\eqref{2ndHe} means that the energy eigenstates (the energy of the adiabatic states) are identical to the energies of the vibrational levels in the diabatic picture, as illustrated in Fig.\ref{schematic}(b). The adiabatic states correspond directly to the laboratory observables.

Vibrational relaxation occurs as a result of escape of system phonons to the environment, modelled as an infinite ensemble of harmonic oscillator modes. The system-bath coupling is then described by
    \begin{equation} \label{H_SB}
        \hat{H}_{B}\color{black}+\hat{H}_{SB} =\sum_{\alpha}\frac{\hat{p}_{\alpha}^2}{2m_{\alpha}}+\frac{1}{2}m_{\alpha}\omega_{\alpha}^2\left(\hat{x}_{\alpha}-\frac{g_{\alpha}\hat{Q}}{2m_{\alpha}\omega_{\alpha}^2}\right)^2, 
    \end{equation}
where $\hat{Q}={(\hat{b}+\hat{b}^{\dagger})}/{\sqrt{2}}$, and $m_{\alpha}$, $\hat{p}_{\alpha}$ and $\hat{x}_{\alpha}$ are the mass, momentum and the coordinate of the environmental harmonic modes,  which correspond to bath phonons\color{black}. The coupling strength $g_{\alpha}$ of the $\alpha$th harmonic oscillator is determined by the spectral density, $J(\omega) = \sum_\alpha{g_{\alpha}^2}{(2m_{\alpha})}^{-1}\omega_{\alpha}\delta(\omega-\omega_{\alpha})$. This model for the surrounding environment is very general and allows in principle even the modelling of non-Markovian system-bath coupling~\cite{Green2019}. Here, however, we work in the Markovian limit and simplify the environment to an overdamped Brownian oscillator profile, $J(\omega)=2\eta{\omega\Lambda}(\omega^2+\Lambda^2)^{-1}$, with bath reorganization energy $\eta$ and dissipation rate $\Lambda$. The overdamped spectral density introduces stochastic, Gaussian fluctuations in the nuclear dynamics, representative of low frequency intermolecular modes from the interaction of the molecule with the solvent. 
The coupling of the system and bath nuclear coordinates leads to vibrational dephasing~\cite{Hamm2011,Ishizaki2008} and dissipation. 
Several different approaches exist for solving the generally computationally demanding equations of motion resulting from Eq.~\eqref{H_tot} and similar open quantum systems~\cite{Kurt2021,Carballeira2021,Gu2021,Alkathiri2022,Avriller2021,Yang2022}. Here we employ the hierarchical equations of motion (HEOM) method~\cite{Tanimura1989,Tanimura2020} in the overdamped limit from Ref.~\cite{Green2019} to evolve the vibronic molecule, equivalent to the Hamiltonian vibration model for a vibronic monomer in Ref.~\cite{Humphries2022}.

The hierarchical equations of motion simulations of the quantum dynamics allow us to numerically compute the correlation functions for the emission of photons and phonons from the molecule. In particular, the correlated emission of photons and phonons is quantified by the normalized second-order correlation function
\begin{equation}
    \label{g2def}
    g_{c_1c_2}^{(2)}(t,\tau)=\frac{\braket{\hat{c}_1(t)\hat{c}_2(t+\tau)\hat{c}_2^{\dagger}(t+\tau)\hat{c}_1^{\dagger}(t)}}{\braket{\hat{c}_1(t)\hat{c}_1^\dagger(t)}\braket{\hat{c}_2(t)\hat{c}_2^\dagger(t)}}.
\end{equation}
When the operators are chosen such that $\hat{c}_{1,2} = \hat{a} = \mu_{eg}\ket{g}\!\bra{e}$, the photon annihilation operator, we obtain the photon-photon correlation function $g_{aa}^{(2)}$, which reflects the joint probability of a photon being emitted at time $t+\tau$ given that a photon was emitted at time $t$. By appropriately choosing $\hat{c}_{1,2}$ from the photon and phonon operators $\hat{a}$ and $\hat{b}$, respectively, we can correspondingly construct the phonon-phonon correlation function $g_{bb}^{(2)}$ and, notably, the photon-phonon and phonon-photon cross-correlation functions $g_{ab}^{(2)}$ and $g_{ba}^{(2)}$ in a manner similar to Refs.~\cite{Moslehi2022,Ma2022,Abo2022,Ban2019,Kalaga2022}.

We now employ the correlation functions~\eqref{g2def} to probe quantum effects~\cite{Fano1961} in the molecular system in its thermal environment. In particular, we determine the presence of anti-bunching~\cite{Paul1982}, defined as $g^{(2)}_{c_1c_2}(t,\tau=0) < g^{(2)}_{c_1c_2}(t,\tau>0)$. This implies that the probability of a second emission event immediately following a first is suppressed. Note that this definition encompasses not only photon-photon or phonon-phonon correlation, but is also generalized~\cite{Ma2022,Moslehi2022} to include cross-correlations where the two emission events consist of one photon and one phonon. Correspondingly, bunching is defined to occur when the probability of simultaneous emission is enhanced, $g^{(2)}_{c_1c_2}(t,\tau=0) > g^{(2)}_{c_1c_2}(t,\tau>0)$.

In the following we assume that all emitted photons and phonons are detected, regardless of scattering directions, e.g., by imagining the system enclosed by a detector and use the quantum regression theorem~\cite{Lax1963,Lax1967,Carmichael1976,Carmichael1999,Mollow1969} to compute second-order correlations as
\begin{equation}
    g_{c_1c_2}^{(2)}(t,\tau)=\frac{\mathrm{Tr}\left[\hat{c}_1^{\dagger}\hat{c}_1\exp(\mathcal{L}\tau)(\hat{c}_2\hat{\rho} \hat{c}_2^{\dagger})\right]}{\mathrm{Tr}(\hat{c}_1\hat{\rho} \hat{c}_1^{\dagger})\mathrm{Tr}(\hat{c}_2\hat{\rho} \hat{c}_2^{\dagger})},\label{second_order}
\end{equation}
where $\hat{\rho}=\hat{\rho}(t)$ is the density matrix at time $t$ and $\mathcal{L}$ is the Liouvillian operator for the time evolution of the system. The molecule is initially equilibrated with its thermal environment, in the absence of the driving field, so that the density matrix of the system is correlated with the bath. This ensures that the vibrational states of the molecule are in the correct Boltzmann distribution. The overdamped hierarchy is then used to evolve the dynamics in the presence of the driving field over several optical cycles to find $\hat{\rho}(t)$. To compute $g_{c_1c_2}^{(2)}(t,\tau)$, one then takes $\hat{c}_2\hat{\rho}(t)\hat{c}_2^\dagger$ as the initial state for a subsequent evolution in $\tau$ taking care to preserve continuity of the driving field.

\begin{figure*}
\centering
  \includegraphics[width=\textwidth]{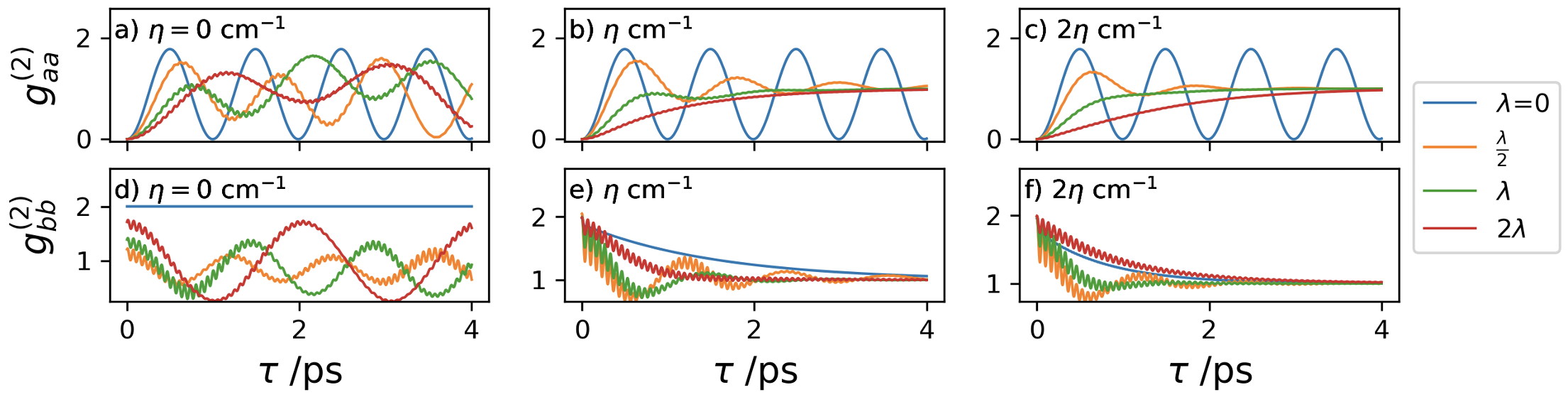}
  \caption{(a)--(c): $g^{(2)}_{aa}(\tau)$ photon-photon correlation function; (d)--(f) $g^{(2)}_{bb}(\tau)$ phonon-phonon correlation function, scanning over bath ($\eta$) and system ($\lambda$) reorganization energies.}\label{g2a_b}
\end{figure*}

%\blue{\textbf{Height: \the\graphicheight}}
%\blue{\textbf{Width: \the\graphicwidth}}
% this is [300/(0.5*AR)]+40 for double column
% h: 128.99805pt w: 510.0pt AR = 3.95354...
% Fig2 words: 192 + 17 caption = 209

Figure~\ref{g2a_b}(a)--(c) show the two-time photon correlation function, $g^{(2)}_{aa}(t,\tau)$, as a function of the photon-photon separation time $\tau$ for the molecular system defined in Eq.~\eqref{H_tot}.
For this and following simulation results, we assume (unless otherwise specified) $\eta = 5 \, \mathrm{cm^{-1}}$, $\Lambda = 200 \,\mathrm{cm^{-1}}$, $\omega_0 = 500 \,\mathrm{cm^{-1}}$, $\Delta = 1.2$ such that $\lambda\approx260 \,\mathrm{cm^{-1}}$, $\omega_{eg} = 10^4 \,\mathrm{cm^{-1}}$, $E_0=10^{7} \,\mathrm{NC^{-1}}$, and $T = 298\,\mathrm{K}$. These parameters ensure that coupling to the bath is weak, and we are operating in the Markovian limit. We truncate the number of vibrational levels at $N=10$, which is sufficiently large for the results to be insensitive to the truncation. These parameters are comparable to real molecules with electronic  and vibrational transition frequencies $\sim10^4$ cm$^{-1}$ and $\sim10^2$ cm$^{-1}$, respectively~\cite{Camargo2015,Lu2020}. The weakly coupled Markovian bath parameters are typical of commonplace non-polar solvents~\cite{Green2019,Humphries2022}.

As population is initialised to a Boltzmann distribution, excitation results in a wavepacket which moves within the harmonic potential~\cite{Nafie1983,Maly2016,Kopec2012}. When the system reorganization energy $\lambda=0$, the effect of the monochromatic laser field is to drive population between the ground vibronic states and the equivalent excited states 
($\ket{g,0}\to\ket{e,0},\ket{g,1}\to\ket{e,1},...$). The resulting Rabi oscillations are reflected in $g^{(2)}_{aa}$ and show in photon anti-bunching. However when $\lambda>0$, VR occurs and the excited state wavepacket population becomes different to that of the ground state. This results in the emergence of a minor oscillation in $g^{(2)}_{aa}$, at the vibrational mode frequency, which implies that the experimentally measurable second-order photon correlation function contains an observable phonon signature  despite the fact that phonons are not directly detected. This signature appears because of the change in Franck-Condon overlap (i.e., the overlap integral of the bound eigenstates of the electronic excited state with the ground state)~\cite{Condon1926}. The Franck-Condon overlap of the fundamental transitions reduces with increasing $\lambda$, increasing the Rabi oscillation period and more population enters the vibronic $\ket{e,0}$ state via VR.

Increased bath reorganization energy $\eta$ damps the Rabi oscillations as phonons dissipate into the bath, leading to the formation of a steady state [Fig.~\ref{g2a_b}(b) and (c)]. We then evaluate $g^{(2)}_{aa}(t,\tau)$ as a function of $\tau$  at a time $t$ after reaching the steady state, in keeping with common convention. When no steady state forms [$\eta=0$, Fig.~\ref{g2a_b}(a)], but Rabi-like oscillations persist indefinitely~\cite{Rabi1937}, the choice of $t$, and therefore the normalisation of $g^{(2)}_{aa}$ as a function of $\tau$ [cf.\ Eq.~\eqref{g2def}] is not obvious~\cite{Glauber1963,Mandel1995,supplement}. We choose $t$ such that the denominator in Eq.~\eqref{g2def} corresponds to its value in the steady state when $\eta>0$.

Figure~\ref{g2a_b}(d)--(f) show the corresponding phonon-phonon correlation, $g^{(2)}_{bb}(\tau)$. When $\lambda=\eta=0$ the population moves resonantly between the ground and excited states and with no VR~\cite{Scully1997}, resulting in a constant phonon-phonon $g^{(2)}_{bb}(\tau)$. Despite lack of phonon dissipation ($\eta=0$), a pronounced oscillation at the mode frequency appears for $\lambda>0$ as the excited state displacement results in a non-stationary population distribution out of thermal equilibrium. For the same reason, $g^{(2)}_{bb}(\tau)$ also tracks the Rabi oscillation when $\lambda > 0$ and phonon bunching is apparent.
Additionally introducing coupling to the environment bath, $\eta>0$, we find a rapid decay of $g^{(2)}_{bb}$ with $\tau$ due to the strong dissipation.

Both the $g^{(2)}_{aa}$ and $g^{(2)}_{bb}$, by definition, have no dependence on the order of detection events since $\tau$ separates the detection of identical particles. In both cases there is a single source of vibrational character. For the $g^{(2)}_{aa}$ this is indirect, from the strong dependence of photons on the vibrational populations, whilst for $g^{(2)}_{bb}$  this is from direct measurement of the phonon number. In both cases, vibrational character is observed as oscillations at the vibrational mode frequency $\omega_0$.
\begin{figure*}
\centering
  \includegraphics[width=\textwidth]{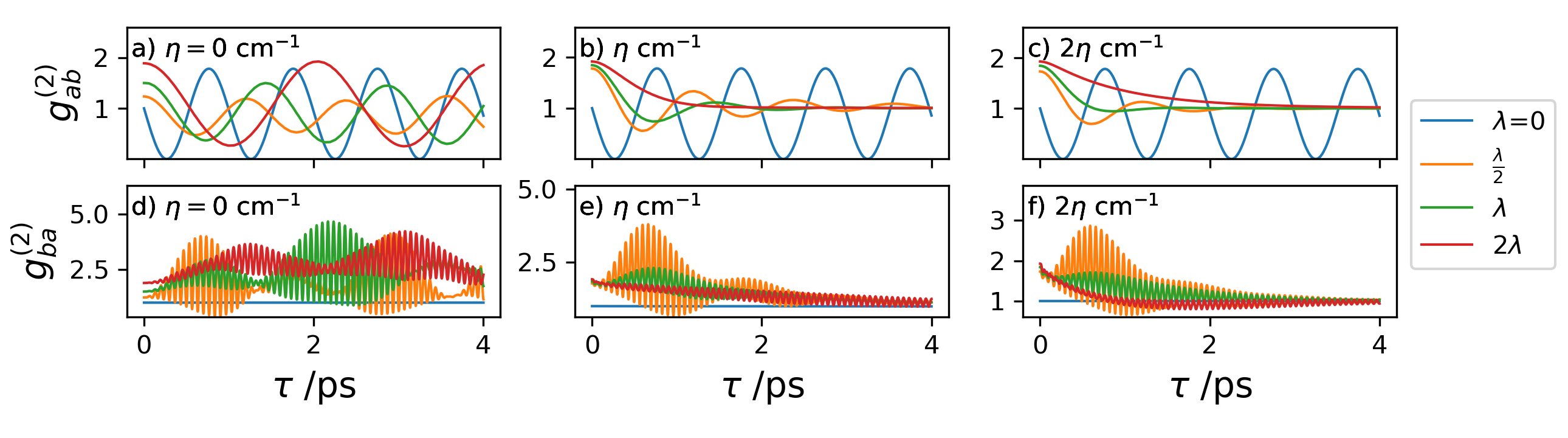}
  \caption{(a)--(c): $g^{(2)}_{ab}(\tau)$ cross-correlation function, (d)--(f): $g^{(2)}_{ba}(\tau)$ cross-correlation function, scanning over bath ($\eta$) and system ($\lambda$) reorganization energies.}\label{g2ab_ba}
\end{figure*}

%\blue{\textbf{Height: \the\graphicheight}}
%\blue{\textbf{Width: \the\graphicwidth}}
% this is [300/(0.5*AR)]+40 for double column
% h: 142.32884pt w: 510.0pt AR = 3.58325...
% Fig3 words: 208 + 17 caption = 225

We can understand the appearance of phonon signatures in the photon correlations from the cross-correlation functions $g^{(2)}_{ab}(\tau)$ and $g^{(2)}_{ba}(\tau)$ (Fig.~\ref{g2ab_ba}), where the order of detection does matter. Specifically, the second detection event determines the dominant character of the cross-correlation function as a function of $\tau$. In $g^{(2)}_{ab}$, the phonon detection is second, and we observe the primary behaviour of the phonon correlation [cf.\ Fig.~\ref{g2a_b} (d)--(f)] with photon correlation-function characteristics superimposed. The first detection event can be thought of as an instantaneous measurement of photon number and  contains no vibrational information. The second detection event---the phonon---occurs a time $\tau$ later, during which vibrational transitions may occur. However, because the fast phonon signatures are very small with respect to the electronic contributions their impact on the excited-state adiabatic population is minimal, i.e., there is no significant minor oscillation~\cite{supplement}. Consequently, neither detection event incurs vibrational character. Figure~\ref{g2ab_ba}(a)--(c) also show phonon bunching, i.e., a photon detection is likely to be immediately followed by another phonon, reflecting the non-equilibrium population distribution following photon emission.

By contrast, the phonon-photon correlation function $g^{(2)}_{ba}(\tau)$ [Fig.~\ref{g2ab_ba}(d)--(f)] corresponds to the observation of a phonon followed by a photon. Since the photon detection is second, characteristics of  $g^{(2)}_{aa}$ [Fig.~\ref{g2a_b}(a)--(c)] dominate, with phonon characteristics superimposed.
The first detection event---the phonon---is an instantaneous measurement of phonon number and thus has intrinsic vibrational character at the molecule mode frequency. However, the second detection event---the photon---also introduces additional vibrational character due to vibrational transitions occurring between the detections. Consequently, there are two sources of vibrational character in $g^{(2)}_{ba}$: 1.\ intrinsically from the first detection event, and 2.\ from phonon effects during the optical cycles leading to the photon emission.

For $\lambda=0$, $g^{(2)}_{ba}(\tau)$ remains at a small, non-zero constant value regardless of the bath coupling. However, increasing the system reorganization energy introduces strong oscillations and the correlation function may drop below the $\lambda=0$ value. This is explained by a large proportion of the wavepacket population undergoing vibrational transitions. This effect persists for small $\tau$ even with strong bath dissipation, but it is destroyed for later $\tau$ by the influence of the environment. Note that similar to $g^{(2)}_{aa}(\tau)$, $g^{(2)}_{ba}(\tau)$ exhibits anti-bunching-like behavior , i.e., a photon is less likely to be emitted directly after a phonon. This is because photon emission at $\omega_{eg}$ from higher vibrational levels is increasingly suppressed for larger $\lambda$ due to decreasing Franck-Condon overlap. This means that phonon emission tends to inhibit subsequent photon emission when $\tau\approx 0$.

Dynamical impact of lattice phonons on quantum-dot emitters embedded in a solid-state system was recently theoretically and experimentally shown to result in a characteristic signature in the photon spectrum~\cite{Brash2019}. System phonons are not defined in the quantum-dot model, but are integral to molecules and the proximate source of the signatures predicted here. Our results suggest that measurements of two-photon correlations could explicitly elucidate differences in the impact of system phonons and the phonon environment.

In conclusion, we have  demonstrated theoretically photon anti-bunching in the fluorescence of a vibronic molecule under continuous laser drive and a thermal environment and that the photon-photon correlations exhibit signatures of the phonon interaction with the bath, suggesting that these are experimentally directly measurable. These appear as oscillations at the system-mode frequency on top of slower modulations associated with the electronic Rabi-like oscillations. 
Theoretically also considering phonon detection and photon-phonon cross-correlation functions, we have shown how vibrational contributions are understood as arising either directly, through phonon detection, or indirectly, through photon detection subsequent to phonon emission. As such, the order of particle detection can dramatically impact the behaviour of the correlation functions, which could in principle be exploited to investigate the phonon impact on photon emission. More immediately, these correlation functions present an opportunity to investigate phonon dynamics indirectly using existing quantum-optical techniques to understand the impact on quantum versus classical processes in molecular systems.  

%\blue{An interesting study by Brash et al.~\cite{Brash2019} looks at photon and phonon dynamics in quantum dots. There are interesting parallels between this study and ours. However, there are some important differences: for example our model explicitly includes both system and bath phonons as opposed to incorporating only environmental phonons. Because we include system phonons we are able to calculate cross-correlation functions and observe explicit system phonon signatures within photon correlations. Whereas in Brash, as a consequence of their dressed states, they are able to observe dynamical impacts of environmental phonons.}

Data used in this publication is available at \cite{Humphries2023}.

\begin{acknowledgements}
The research presented in this paper was carried out on the High Performance Computing Cluster supported by the Research and Specialist Computing Support service at the University of East Anglia. B.S.H. thanks the Faculty of Science, University of East Anglia, for studentship funding. G.A.J. and D.G., and M.O.B. acknowledge support from the Engineering and Physical Sciences Research Council under Awards No. EP/V00817X/1 and EP/V03832X/1, respectively. For the purpose of open access, the authors have applied a Creative Commons Attribution (CC~BY) licence to any Author Accepted Manuscript version arising.
The authors express their gratitude to Dr. Kayn Forbes for discussion.
\end{acknowledgements}

%\bibliography{MS_bib}

\begin{thebibliography}{63}%
\makeatletter
\providecommand \@ifxundefined [1]{%
 \@ifx{#1\undefined}
}%
\providecommand \@ifnum [1]{%
 \ifnum #1\expandafter \@firstoftwo
 \else \expandafter \@secondoftwo
 \fi
}%
\providecommand \@ifx [1]{%
 \ifx #1\expandafter \@firstoftwo
 \else \expandafter \@secondoftwo
 \fi
}%
\providecommand \natexlab [1]{#1}%
\providecommand \enquote  [1]{``#1''}%
\providecommand \bibnamefont  [1]{#1}%
\providecommand \bibfnamefont [1]{#1}%
\providecommand \citenamefont [1]{#1}%
\providecommand \href@noop [0]{\@secondoftwo}%
\providecommand \href [0]{\begingroup \@sanitize@url \@href}%
\providecommand \@href[1]{\@@startlink{#1}\@@href}%
\providecommand \@@href[1]{\endgroup#1\@@endlink}%
\providecommand \@sanitize@url [0]{\catcode `\\12\catcode `\$12\catcode
  `\&12\catcode `\#12\catcode `\^12\catcode `\_12\catcode `\%12\relax}%
\providecommand \@@startlink[1]{}%
\providecommand \@@endlink[0]{}%
\providecommand \url  [0]{\begingroup\@sanitize@url \@url }%
\providecommand \@url [1]{\endgroup\@href {#1}{\urlprefix }}%
\providecommand \urlprefix  [0]{URL }%
\providecommand \Eprint [0]{\href }%
\providecommand \doibase [0]{http://dx.doi.org/}%
\providecommand \selectlanguage [0]{\@gobble}%
\providecommand \bibinfo  [0]{\@secondoftwo}%
\providecommand \bibfield  [0]{\@secondoftwo}%
\providecommand \translation [1]{[#1]}%
\providecommand \BibitemOpen [0]{}%
\providecommand \bibitemStop [0]{}%
\providecommand \bibitemNoStop [0]{.\EOS\space}%
\providecommand \EOS [0]{\spacefactor3000\relax}%
\providecommand \BibitemShut  [1]{\csname bibitem#1\endcsname}%
\let\auto@bib@innerbib\@empty
%</preamble>
\bibitem [{\citenamefont {Kimble}\ \emph {et~al.}(1977)\citenamefont {Kimble},
  \citenamefont {Dagenais},\ and\ \citenamefont {Mandel}}]{Kimble1977}%
  \BibitemOpen
  \bibfield  {author} {\bibinfo {author} {\bibfnamefont {H.~J.}\ \bibnamefont
  {Kimble}}, \bibinfo {author} {\bibfnamefont {M.}~\bibnamefont {Dagenais}}, \
  and\ \bibinfo {author} {\bibfnamefont {L.}~\bibnamefont {Mandel}},\
  }\bibfield  {title} {\enquote {\bibinfo {title} {Photon antibunching in
  resonance fluorescence},}\ }\href {\doibase 10.1103/PhysRevLett.39.691}
  {\bibfield  {journal} {\bibinfo  {journal} {Phys. Rev. Lett.}\ }\textbf
  {\bibinfo {volume} {39}},\ \bibinfo {pages} {691} (\bibinfo {year}
  {1977})}\BibitemShut {NoStop}%
\bibitem [{\citenamefont {Kimble}\ \emph {et~al.}(1978)\citenamefont {Kimble},
  \citenamefont {Dagenais},\ and\ \citenamefont {Mandel}}]{Kimble1978}%
  \BibitemOpen
  \bibfield  {author} {\bibinfo {author} {\bibfnamefont {H.~J.}\ \bibnamefont
  {Kimble}}, \bibinfo {author} {\bibfnamefont {M.}~\bibnamefont {Dagenais}}, \
  and\ \bibinfo {author} {\bibfnamefont {L.}~\bibnamefont {Mandel}},\
  }\bibfield  {title} {\enquote {\bibinfo {title} {Multiatom and transit-time
  effects on photon-correlation measurements in resonance fluorescence},}\
  }\href {\doibase 10.1103/PhysRevA.18.201} {\bibfield  {journal} {\bibinfo
  {journal} {Phys. Rev. A}\ }\textbf {\bibinfo {volume} {18}},\ \bibinfo
  {pages} {201} (\bibinfo {year} {1978})}\BibitemShut {NoStop}%
\bibitem [{\citenamefont {Walls}(1979)}]{Walls1979}%
  \BibitemOpen
  \bibfield  {author} {\bibinfo {author} {\bibfnamefont {D.~F.}\ \bibnamefont
  {Walls}},\ }\bibfield  {title} {\enquote {\bibinfo {title} {{Evidence for the
  quantum nature of light}},}\ }\href {\doibase 10.1038/280451a0} {\bibfield
  {journal} {\bibinfo  {journal} {Nature}\ }\textbf {\bibinfo {volume} {280}},\
  \bibinfo {pages} {451} (\bibinfo {year} {1979})}\BibitemShut {NoStop}%
\bibitem [{\citenamefont {Paul}(1982)}]{Paul1982}%
  \BibitemOpen
  \bibfield  {author} {\bibinfo {author} {\bibfnamefont {H.}~\bibnamefont
  {Paul}},\ }\bibfield  {title} {\enquote {\bibinfo {title} {Photon
  antibunching},}\ }\href {\doibase 10.1103/RevModPhys.54.1061} {\bibfield
  {journal} {\bibinfo  {journal} {Rev. Mod. Phys.}\ }\textbf {\bibinfo {volume}
  {54}},\ \bibinfo {pages} {1061} (\bibinfo {year} {1982})}\BibitemShut
  {NoStop}%
\bibitem [{\citenamefont {Carmichael}\ and\ \citenamefont
  {Walls}(1976)}]{Carmichael1976}%
  \BibitemOpen
  \bibfield  {author} {\bibinfo {author} {\bibfnamefont {H.~J.}\ \bibnamefont
  {Carmichael}}\ and\ \bibinfo {author} {\bibfnamefont {D.~F.}\ \bibnamefont
  {Walls}},\ }\bibfield  {title} {\enquote {\bibinfo {title} {{A
  quantum-mechanical master equation treatment of the dynamical Stark
  effect}},}\ }\href {\doibase 10.1088/0022-3700/9/8/007} {\bibfield  {journal}
  {\bibinfo  {journal} {J. Phys. B At. Mol. Phys.}\ }\textbf {\bibinfo {volume}
  {9}},\ \bibinfo {pages} {1199} (\bibinfo {year} {1976})}\BibitemShut
  {NoStop}%
\bibitem [{\citenamefont {Zheng}\ \emph {et~al.}(2012)\citenamefont {Zheng},
  \citenamefont {Gauthier},\ and\ \citenamefont {Baranger}}]{Zheng2012}%
  \BibitemOpen
  \bibfield  {author} {\bibinfo {author} {\bibfnamefont {Huaixiu}\ \bibnamefont
  {Zheng}}, \bibinfo {author} {\bibfnamefont {Daniel~J.}\ \bibnamefont
  {Gauthier}}, \ and\ \bibinfo {author} {\bibfnamefont {Harold~U.}\
  \bibnamefont {Baranger}},\ }\bibfield  {title} {\enquote {\bibinfo {title}
  {Strongly correlated photons generated by coupling a three- or four-level
  system to a waveguide},}\ }\href {\doibase 10.1103/PhysRevA.85.043832}
  {\bibfield  {journal} {\bibinfo  {journal} {Phys. Rev. A}\ }\textbf {\bibinfo
  {volume} {85}},\ \bibinfo {pages} {043832} (\bibinfo {year}
  {2012})}\BibitemShut {NoStop}%
\bibitem [{\citenamefont {Avriller}\ \emph {et~al.}(2021)\citenamefont
  {Avriller}, \citenamefont {Schaeverbeke}, \citenamefont {Frederiksen},\ and\
  \citenamefont {Pistolesi}}]{Avriller2021}%
  \BibitemOpen
  \bibfield  {author} {\bibinfo {author} {\bibfnamefont {R.}~\bibnamefont
  {Avriller}}, \bibinfo {author} {\bibfnamefont {Q.}~\bibnamefont
  {Schaeverbeke}}, \bibinfo {author} {\bibfnamefont {T.}~\bibnamefont
  {Frederiksen}}, \ and\ \bibinfo {author} {\bibfnamefont {F.}~\bibnamefont
  {Pistolesi}},\ }\bibfield  {title} {\enquote {\bibinfo {title}
  {Photon-emission statistics induced by electron tunneling in plasmonic
  nanojunctions},}\ }\href {\doibase 10.1103/PhysRevB.104.L241403} {\bibfield
  {journal} {\bibinfo  {journal} {Phys. Rev. B}\ }\textbf {\bibinfo {volume}
  {104}},\ \bibinfo {pages} {L241403} (\bibinfo {year} {2021})}\BibitemShut
  {NoStop}%
\bibitem [{\citenamefont {Ishida}\ and\ \citenamefont
  {Nasu}(2012)}]{Ishida2012}%
  \BibitemOpen
  \bibfield  {author} {\bibinfo {author} {\bibfnamefont {Kunio}\ \bibnamefont
  {Ishida}}\ and\ \bibinfo {author} {\bibfnamefont {Keiichiro}\ \bibnamefont
  {Nasu}},\ }\bibfield  {title} {\enquote {\bibinfo {title} {Theoretical study
  of transient x-ray speckles in photoinduced nucleation},}\ }\href {\doibase 10.1143/JPSJ.81.063708} {\bibfield  {journal} {\bibinfo  {journal} {J.
  Phys. Soc. Jpn}\ }\textbf {\bibinfo {volume} {81}},\
  \bibinfo {pages} {063708} (\bibinfo {year} {2012})}\ 
  %\Eprint
  %{http://arxiv.org/abs/https://doi.org/10.1143/JPSJ.81.063708}
  %{https://doi.org/10.1143/JPSJ.81.063708} 
  \BibitemShut {NoStop}%
\bibitem [{\citenamefont {Imamo\ifmmode~\bar{g}\else \={g}\fi{}lu}\ \emph
  {et~al.}(1997)\citenamefont {Imamo\ifmmode~\bar{g}\else \={g}\fi{}lu},
  \citenamefont {Schmidt}, \citenamefont {Woods},\ and\ \citenamefont
  {Deutsch}}]{Imamoglu1997}%
  \BibitemOpen
  \bibfield  {author} {\bibinfo {author} {\bibfnamefont {A.}~\bibnamefont
  {Imamo\ifmmode~\bar{g}\else \={g}\fi{}lu}}, \bibinfo {author} {\bibfnamefont
  {H.}~\bibnamefont {Schmidt}}, \bibinfo {author} {\bibfnamefont
  {G.}~\bibnamefont {Woods}}, \ and\ \bibinfo {author} {\bibfnamefont
  {M.}~\bibnamefont {Deutsch}},\ }\bibfield  {title} {\enquote {\bibinfo
  {title} {Strongly interacting photons in a nonlinear cavity},}\ }\href
  {\doibase 10.1103/PhysRevLett.79.1467} {\bibfield  {journal} {\bibinfo
  {journal} {Phys. Rev. Lett.}\ }\textbf {\bibinfo {volume} {79}},\ \bibinfo
  {pages} {1467} (\bibinfo {year} {1997})}\BibitemShut {NoStop}%
\bibitem [{\citenamefont {Birnbaum}\ \emph {et~al.}(2005)\citenamefont
  {Birnbaum}, \citenamefont {Boca}, \citenamefont {Miller}, \citenamefont
  {Boozer}, \citenamefont {Northup},\ and\ \citenamefont
  {Kimble}}]{Birnbaum2005}%
  \BibitemOpen
  \bibfield  {author} {\bibinfo {author} {\bibfnamefont {K.~M.}\ \bibnamefont
  {Birnbaum}}, \bibinfo {author} {\bibfnamefont {A.}~\bibnamefont {Boca}},
  \bibinfo {author} {\bibfnamefont {R.}~\bibnamefont {Miller}}, \bibinfo
  {author} {\bibfnamefont {A.~D.}\ \bibnamefont {Boozer}}, \bibinfo {author}
  {\bibfnamefont {T.~E.}\ \bibnamefont {Northup}}, \ and\ \bibinfo {author}
  {\bibfnamefont {H.~J.}\ \bibnamefont {Kimble}},\ }\bibfield  {title}
  {\enquote {\bibinfo {title} {{Photon blockade in an optical cavity with one
  trapped atom}},}\ }\href {\doibase 10.1038/nature03804} {\bibfield  {journal}
  {\bibinfo  {journal} {Nature}\ }\textbf {\bibinfo {volume} {436}},\ \bibinfo
  {pages} {87} (\bibinfo {year} {2005})}
  %\ \Eprint
  %{http://arxiv.org/abs/0507065v1} {arXiv:0507065v1 [arXiv:quant-ph]}
  \BibitemShut {NoStop}%
\bibitem [{\citenamefont {Lang}\ \emph {et~al.}(2011)\citenamefont {Lang},
  \citenamefont {Bozyigit}, \citenamefont {Eichler}, \citenamefont {Steffen},
  \citenamefont {Fink}, \citenamefont {Abdumalikov}, \citenamefont {Baur},
  \citenamefont {Filipp}, \citenamefont {da~Silva}, \citenamefont {Blais},\
  and\ \citenamefont {Wallraff}}]{Lang2011}%
  \BibitemOpen
  \bibfield  {author} {\bibinfo {author} {\bibfnamefont {C.}~\bibnamefont
  {Lang}}, \bibinfo {author} {\bibfnamefont {D.}~\bibnamefont {Bozyigit}},
  \bibinfo {author} {\bibfnamefont {C.}~\bibnamefont {Eichler}}, \bibinfo
  {author} {\bibfnamefont {L.}~\bibnamefont {Steffen}}, \bibinfo {author}
  {\bibfnamefont {J.~M.}\ \bibnamefont {Fink}}, \bibinfo {author}
  {\bibfnamefont {A.~A.}\ \bibnamefont {Abdumalikov}}, \bibinfo {author}
  {\bibfnamefont {M.}~\bibnamefont {Baur}}, \bibinfo {author} {\bibfnamefont
  {S.}~\bibnamefont {Filipp}}, \bibinfo {author} {\bibfnamefont {M.~P.}\
  \bibnamefont {da~Silva}}, \bibinfo {author} {\bibfnamefont {A.}~\bibnamefont
  {Blais}}, \ and\ \bibinfo {author} {\bibfnamefont {A.}~\bibnamefont
  {Wallraff}},\ }\bibfield  {title} {\enquote {\bibinfo {title} {Observation of
  resonant photon blockade at microwave frequencies using correlation function
  measurements},}\ }\href {\doibase 10.1103/PhysRevLett.106.243601} {\bibfield
  {journal} {\bibinfo  {journal} {Phys. Rev. Lett.}\ }\textbf {\bibinfo
  {volume} {106}},\ \bibinfo {pages} {243601} (\bibinfo {year}
  {2011})}\BibitemShut {NoStop}%
\bibitem [{\citenamefont {Deng}\ \emph {et~al.}(2017)\citenamefont {Deng},
  \citenamefont {Li},\ and\ \citenamefont {Qin}}]{Deng2017}%
  \BibitemOpen
  \bibfield  {author} {\bibinfo {author} {\bibfnamefont {Wen-Wu}\ \bibnamefont
  {Deng}}, \bibinfo {author} {\bibfnamefont {Gao-Xiang}\ \bibnamefont {Li}}, \
  and\ \bibinfo {author} {\bibfnamefont {Hong}\ \bibnamefont {Qin}},\
  }\bibfield  {title} {\enquote {\bibinfo {title} {Photon blockade via quantum
  interference in a strong coupling qubit-cavity system},}\ }\href {\doibase 10.1364/OE.25.006767} {\bibfield  {journal} {\bibinfo  {journal} {Opt.
  Express}\ }\textbf {\bibinfo {volume} {25}},\ \bibinfo {pages} {6767}
  (\bibinfo {year} {2017})}\BibitemShut {NoStop}%
\bibitem [{\citenamefont {Gu}\ and\ \citenamefont {Mukamel}(2022)}]{Gu2021}%
  \BibitemOpen
  \bibfield  {author} {\bibinfo {author} {\bibfnamefont {Bing}\ \bibnamefont
  {Gu}}\ and\ \bibinfo {author} {\bibfnamefont {Shaul}\ \bibnamefont
  {Mukamel}},\ }\bibfield  {title} {\enquote {\bibinfo {title} {Photon
  correlation signals in coupled-cavity polaritons created by entangled
  light},}\ }\href {\doibase 10.1021/acsphotonics.1c01755} {\bibfield
  {journal} {\bibinfo  {journal} {ACS Photonics}\ }\textbf {\bibinfo {volume}
  {9}},\ \bibinfo {pages} {938} (\bibinfo {year} {2022})}
  %\ \Eprint
  %{http://arxiv.org/abs/https://doi.org/10.1021/acsphotonics.1c0175}
  %{https://doi.org/10.1021/acsphotonics.1c0175} 
  \BibitemShut {NoStop}%
\bibitem [{\citenamefont {Guo}\ \emph {et~al.}(2022)\citenamefont {Guo},
  \citenamefont {Zou}, \citenamefont {Huang},\ and\ \citenamefont
  {Liao}}]{Guo2022}%
  \BibitemOpen
  \bibfield  {author} {\bibinfo {author} {\bibfnamefont {Ya-Ting}\ \bibnamefont
  {Guo}}, \bibinfo {author} {\bibfnamefont {Fen}\ \bibnamefont {Zou}}, \bibinfo
  {author} {\bibfnamefont {Jin-Feng}\ \bibnamefont {Huang}}, \ and\ \bibinfo
  {author} {\bibfnamefont {Jie-Qiao}\ \bibnamefont {Liao}},\ }\bibfield
  {title} {\enquote {\bibinfo {title} {Retrieval of photon blockade effect in
  the dispersive jaynes-cummings model},}\ }\href {\doibase 10.1103/PhysRevA.105.013705} {\bibfield  {journal} {\bibinfo  {journal}
  {Phys. Rev. A}\ }\textbf {\bibinfo {volume} {105}},\ \bibinfo {pages}
  {013705} (\bibinfo {year} {2022})}\BibitemShut {NoStop}%
\bibitem [{\citenamefont {Ridolfo}\ \emph {et~al.}(2012)\citenamefont
  {Ridolfo}, \citenamefont {Leib}, \citenamefont {Savasta},\ and\ \citenamefont
  {Hartmann}}]{Ridolfo2012}%
  \BibitemOpen
  \bibfield  {author} {\bibinfo {author} {\bibfnamefont {A.}~\bibnamefont
  {Ridolfo}}, \bibinfo {author} {\bibfnamefont {M.}~\bibnamefont {Leib}},
  \bibinfo {author} {\bibfnamefont {S.}~\bibnamefont {Savasta}}, \ and\
  \bibinfo {author} {\bibfnamefont {M.~J.}\ \bibnamefont {Hartmann}},\
  }\bibfield  {title} {\enquote {\bibinfo {title} {Photon blockade in the
  ultrastrong coupling regime},}\ }\href {\doibase 10.1103/PhysRevLett.109.193602} {\bibfield  {journal} {\bibinfo  {journal}
  {Phys. Rev. Lett.}\ }\textbf {\bibinfo {volume} {109}},\ \bibinfo {pages}
  {193602} (\bibinfo {year} {2012})}\BibitemShut {NoStop}%
\bibitem [{\citenamefont {Cidrim}\ \emph {et~al.}(2020)\citenamefont {Cidrim},
  \citenamefont {do~Espirito~Santo}, \citenamefont {Schachenmayer},
  \citenamefont {Kaiser},\ and\ \citenamefont {Bachelard}}]{Cidrim2020}%
  \BibitemOpen
  \bibfield  {author} {\bibinfo {author} {\bibfnamefont {A.}~\bibnamefont
  {Cidrim}}, \bibinfo {author} {\bibfnamefont {T.~S.}\ \bibnamefont
  {do~Espirito~Santo}}, \bibinfo {author} {\bibfnamefont {J.}~\bibnamefont
  {Schachenmayer}}, \bibinfo {author} {\bibfnamefont {R.}~\bibnamefont
  {Kaiser}}, \ and\ \bibinfo {author} {\bibfnamefont {R.}~\bibnamefont
  {Bachelard}},\ }\bibfield  {title} {\enquote {\bibinfo {title} {Photon
  blockade with ground-state neutral atoms},}\ }\href {\doibase 10.1103/PhysRevLett.125.073601} {\bibfield  {journal} {\bibinfo  {journal}
  {Phys. Rev. Lett.}\ }\textbf {\bibinfo {volume} {125}},\ \bibinfo {pages}
  {073601} (\bibinfo {year} {2020})}\BibitemShut {NoStop}%
\bibitem [{\citenamefont {Williamson}\ \emph {et~al.}(2020)\citenamefont
  {Williamson}, \citenamefont {Borgh},\ and\ \citenamefont
  {Ruostekoski}}]{Williamson2020a}%
  \BibitemOpen
  \bibfield  {author} {\bibinfo {author} {\bibfnamefont {L.~A.}\ \bibnamefont
  {Williamson}}, \bibinfo {author} {\bibfnamefont {M.~O.}\ \bibnamefont
  {Borgh}}, \ and\ \bibinfo {author} {\bibfnamefont {J.}~\bibnamefont
  {Ruostekoski}},\ }\bibfield  {title} {\enquote {\bibinfo {title} {Superatom
  picture of collective nonclassical light emission and dipole blockade in atom
  arrays},}\ }\href {\doibase 10.1103/PhysRevLett.125.073602} {\bibfield
  {journal} {\bibinfo  {journal} {Phys. Rev. Lett.}\ }\textbf {\bibinfo
  {volume} {125}},\ \bibinfo {pages} {073602} (\bibinfo {year}
  {2020})}\BibitemShut {NoStop}%
\bibitem [{\citenamefont {Chen}\ \emph {et~al.}(2022)\citenamefont {Chen},
  \citenamefont {Tang}, \citenamefont {Tang}, \citenamefont {Wu},\ and\
  \citenamefont {Xia}}]{Chen2022}%
  \BibitemOpen
  \bibfield  {author} {\bibinfo {author} {\bibfnamefont {Mingyuan}\
  \bibnamefont {Chen}}, \bibinfo {author} {\bibfnamefont {Jiangshan}\
  \bibnamefont {Tang}}, \bibinfo {author} {\bibfnamefont {Lei}\ \bibnamefont
  {Tang}}, \bibinfo {author} {\bibfnamefont {Haodong}\ \bibnamefont {Wu}}, \
  and\ \bibinfo {author} {\bibfnamefont {Keyu}\ \bibnamefont {Xia}},\
  }\bibfield  {title} {\enquote {\bibinfo {title} {Photon blockade and
  single-photon generation with multiple quantum emitters},}\ }\href {\doibase 10.1103/PhysRevResearch.4.033083} {\bibfield  {journal} {\bibinfo  {journal}
  {Phys. Rev. Res.}\ }\textbf {\bibinfo {volume} {4}},\ \bibinfo {pages}
  {033083} (\bibinfo {year} {2022})}\BibitemShut {NoStop}%
\bibitem [{\citenamefont {Feng}\ and\ \citenamefont {Gong}(2021)}]{Feng2021}%
  \BibitemOpen
  \bibfield  {author} {\bibinfo {author} {\bibfnamefont {Ling-Juan}\
  \bibnamefont {Feng}}\ and\ \bibinfo {author} {\bibfnamefont {Shang-Qing}\
  \bibnamefont {Gong}},\ }\bibfield  {title} {\enquote {\bibinfo {title}
  {Two-photon blockade generated and enhanced by mechanical squeezing},}\
  }\href {\doibase 10.1103/PhysRevA.103.043509} {\bibfield  {journal} {\bibinfo
   {journal} {Phys. Rev. A}\ }\textbf {\bibinfo {volume} {103}},\ \bibinfo
  {pages} {043509} (\bibinfo {year} {2021})}\BibitemShut {NoStop}%
\bibitem [{\citenamefont {{Hassani Nadiki}}\ and\ \citenamefont
  {{Tavassoly}}(2022)}]{HassaniNadiki2022}%
  \BibitemOpen
  \bibfield  {author} {\bibinfo {author} {\bibfnamefont {M.}~\bibnamefont
  {{Hassani Nadiki}}}\ and\ \bibinfo {author} {\bibfnamefont {M.~K.}\
  \bibnamefont {{Tavassoly}}},\ }\bibfield  {title} {\enquote {\bibinfo {title}
  {{Phonon blockade in a system consisting of two optomechanical cavities with
  quadratic cavity-membrane coupling and phonon hopping}},}\ }\href {\doibase 10.1140/epjd/s10053-022-00384-9} {\bibfield  {journal} {\bibinfo  {journal}
  {Eur. Phys. J. D}\ }\textbf {\bibinfo {volume} {76}},\ \bibinfo
  {eid} {58} (\bibinfo {year} {2022})}\BibitemShut {NoStop}%
\bibitem [{\citenamefont {Li}\ \emph {et~al.}(2022)\citenamefont {Li},
  \citenamefont {Jin}, \citenamefont {Yin},\ and\ \citenamefont
  {Chen}}]{Li2022}%
  \BibitemOpen
  \bibfield  {author} {\bibinfo {author} {\bibfnamefont {Zi-Yuan}\ \bibnamefont
  {Li}}, \bibinfo {author} {\bibfnamefont {Guang-Ri}\ \bibnamefont {Jin}},
  \bibinfo {author} {\bibfnamefont {Tai-Shuang}\ \bibnamefont {Yin}}, \ and\
  \bibinfo {author} {\bibfnamefont {Aixi}\ \bibnamefont {Chen}},\ }\bibfield
  {title} {\enquote {\bibinfo {title} {Two-phonon blockade in quadratically
  coupled optomechanical systems},}\ }\href {\doibase 10.3390/photonics9020070}
  {\bibfield  {journal} {\bibinfo  {journal} {Photonics}\ }\textbf {\bibinfo
  {volume} {9}}, \bibinfo {pages}
  {70}
  (\bibinfo {year} {2022})}\
  %10.3390/photonics9020070}
  \BibitemShut {NoStop}%
\bibitem [{\citenamefont {Wang}\ \emph {et~al.}(2022)\citenamefont {Wang},
  \citenamefont {Wu}, \citenamefont {Han}, \citenamefont {Xia}, \citenamefont
  {Jiang},\ and\ \citenamefont {Song}}]{Wang2022}%
  \BibitemOpen
  \bibfield  {author} {\bibinfo {author} {\bibfnamefont {Yan}\ \bibnamefont
  {Wang}}, \bibinfo {author} {\bibfnamefont {Jin~Lei}\ \bibnamefont {Wu}},
  \bibinfo {author} {\bibfnamefont {Jin~Xuan}\ \bibnamefont {Han}}, \bibinfo
  {author} {\bibfnamefont {Yan}\ \bibnamefont {Xia}}, \bibinfo {author}
  {\bibfnamefont {Yong~Yuan}\ \bibnamefont {Jiang}}, \ and\ \bibinfo {author}
  {\bibfnamefont {Jie}\ \bibnamefont {Song}},\ }\bibfield  {title} {\enquote
  {\bibinfo {title} {{Enhanced Phonon Blockade in a Weakly Coupled Hybrid
  System via Mechanical Parametric Amplification}},}\ }\href {\doibase 10.1103/PhysRevApplied.17.024009} {\bibfield  {journal} {\bibinfo  {journal}
  {Phys. Rev. Appl.}\ }\textbf {\bibinfo {volume} {17}},\ \bibinfo {pages}
  {024009} (\bibinfo {year} {2022})}
  %\ \Eprint {http://arxiv.org/abs/2112.08562}
  %{arXiv:2112.08562} 
  \BibitemShut {NoStop}%
\bibitem [{\citenamefont {Yin}\ \emph {et~al.}(2022)\citenamefont {Yin},
  \citenamefont {Jin},\ and\ \citenamefont {Chen}}]{Yin2022}%
  \BibitemOpen
  \bibfield  {author} {\bibinfo {author} {\bibfnamefont {Tai-Shuang}\
  \bibnamefont {Yin}}, \bibinfo {author} {\bibfnamefont {Guang-Ri}\
  \bibnamefont {Jin}}, \ and\ \bibinfo {author} {\bibfnamefont {Aixi}\
  \bibnamefont {Chen}},\ }\bibfield  {title} {\enquote {\bibinfo {title}
  {Enhanced phonon antibunching in a circuit quantum acoustodynamical system
  containing two surface acoustic wave resonators},}\ }\href {\doibase 10.3390/mi13040591} {\bibfield  {journal} {\bibinfo  {journal}
  {Micromachines}\ }\textbf {\bibinfo {volume} {13}}, \bibinfo {pages} {591} (\bibinfo {year} {2022})}\
  %10.3390/mi13040591}
  \BibitemShut {NoStop}%
\bibitem [{\citenamefont {{Moslehi}}\ \emph {et~al.}(2022)\citenamefont
  {{Moslehi}}, \citenamefont {{Baghshahi}}, \citenamefont {{Faghihi}},\ and\
  \citenamefont {{Mirafzali}}}]{Moslehi2022}%
  \BibitemOpen
  \bibfield  {author} {\bibinfo {author} {\bibfnamefont {Mahboobeh}\
  \bibnamefont {{Moslehi}}}, \bibinfo {author} {\bibfnamefont {Hamid~Reza}\
  \bibnamefont {{Baghshahi}}}, \bibinfo {author} {\bibfnamefont
  {Mohammad~Javad}\ \bibnamefont {{Faghihi}}}, \ and\ \bibinfo {author}
  {\bibfnamefont {Sayyed~Yahya}\ \bibnamefont {{Mirafzali}}},\ }\bibfield
  {title} {\enquote {\bibinfo {title} {{Photon and magnon blockade induced by
  optomagnonic microcavity}},}\ }\href {\doibase 10.1140/epjp/s13360-022-02937-8} {\bibfield  {journal} {\bibinfo  {journal}
  {Eur. Phys. J. Plus}\ }\textbf {\bibinfo {volume} {137}},\
  \bibinfo {eid} {777} (\bibinfo {year} {2022})}\BibitemShut {NoStop}%
\bibitem [{\citenamefont {Ma}\ \emph {et~al.}(2022)\citenamefont {Ma},
  \citenamefont {Horoshko}, \citenamefont {Yu},\ and\ \citenamefont
  {Kilin}}]{Ma2022}%
  \BibitemOpen
  \bibfield  {author} {\bibinfo {author} {\bibfnamefont {Ting-ting}\
  \bibnamefont {Ma}}, \bibinfo {author} {\bibfnamefont {Dmitri~B.}\
  \bibnamefont {Horoshko}}, \bibinfo {author} {\bibfnamefont {Chang-shui}\
  \bibnamefont {Yu}}, \ and\ \bibinfo {author} {\bibfnamefont {Sergei~Ya.}\
  \bibnamefont {Kilin}},\ }\bibfield  {title} {\enquote {\bibinfo {title}
  {Photon and phonon statistics in a qubit-plasmon-phonon ultrastrong-coupling
  system},}\ }\href {\doibase 10.1103/PhysRevA.105.053718} {\bibfield
  {journal} {\bibinfo  {journal} {Phys. Rev. A}\ }\textbf {\bibinfo {volume}
  {105}},\ \bibinfo {pages} {053718} (\bibinfo {year} {2022})}\BibitemShut
  {NoStop}%
\bibitem [{\citenamefont {Scholes}\ \emph {et~al.}(2017)\citenamefont
  {Scholes}, \citenamefont {Fleming}, \citenamefont {Chen}, \citenamefont
  {Aspuru-Guzik}, \citenamefont {Buchleitner}, \citenamefont {Coker},
  \citenamefont {Engel}, \citenamefont {{Van Grondelle}}, \citenamefont
  {Ishizaki}, \citenamefont {Jonas}, \citenamefont {Lundeen}, \citenamefont
  {McCusker}, \citenamefont {Mukamel}, \citenamefont {Ogilvie}, \citenamefont
  {Olaya-Castro}, \citenamefont {Ratner}, \citenamefont {Spano}, \citenamefont
  {Whaley},\ and\ \citenamefont {Zhu}}]{Scholes2017}%
  \BibitemOpen
  \bibfield  {author} {\bibinfo {author} {\bibfnamefont {Gregory~D.}\
  \bibnamefont {Scholes}}, \bibinfo {author} {\bibfnamefont {Graham~R.}\
  \bibnamefont {Fleming}}, \bibinfo {author} {\bibfnamefont {Lin~X.}\
  \bibnamefont {Chen}}, \bibinfo {author} {\bibfnamefont {Al{\'{a}}n}\
  \bibnamefont {Aspuru-Guzik}}, \bibinfo {author} {\bibfnamefont {Andreas}\
  \bibnamefont {Buchleitner}}, \bibinfo {author} {\bibfnamefont {David~F.}\
  \bibnamefont {Coker}}, \bibinfo {author} {\bibfnamefont {Gregory~S.}\
  \bibnamefont {Engel}}, \bibinfo {author} {\bibfnamefont {Rienk}\ \bibnamefont
  {{Van Grondelle}}}, \bibinfo {author} {\bibfnamefont {Akihito}\ \bibnamefont
  {Ishizaki}}, \bibinfo {author} {\bibfnamefont {David~M.}\ \bibnamefont
  {Jonas}}, \bibinfo {author} {\bibfnamefont {Jeff~S.}\ \bibnamefont
  {Lundeen}}, \bibinfo {author} {\bibfnamefont {James~K.}\ \bibnamefont
  {McCusker}}, \bibinfo {author} {\bibfnamefont {Shaul}\ \bibnamefont
  {Mukamel}}, \bibinfo {author} {\bibfnamefont {Jennifer~P.}\ \bibnamefont
  {Ogilvie}}, \bibinfo {author} {\bibfnamefont {Alexandra}\ \bibnamefont
  {Olaya-Castro}}, \bibinfo {author} {\bibfnamefont {Mark~A.}\ \bibnamefont
  {Ratner}}, \bibinfo {author} {\bibfnamefont {Frank~C.}\ \bibnamefont
  {Spano}}, \bibinfo {author} {\bibfnamefont {K.~Birgitta}\ \bibnamefont
  {Whaley}}, \ and\ \bibinfo {author} {\bibfnamefont {Xiaoyang}\ \bibnamefont
  {Zhu}},\ }\bibfield  {title} {\enquote {\bibinfo {title} {{Using coherence to
  enhance function in chemical and biophysical systems}},}\ }\href {\doibase 10.1038/nature21425} {\bibfield  {journal} {\bibinfo  {journal} {Nature}\
  }\textbf {\bibinfo {volume} {543}},\ \bibinfo {pages} {647} (\bibinfo
  {year} {2017})}\BibitemShut {NoStop}%
\bibitem [{\citenamefont {Jones}\ and\ \citenamefont
  {Bradshaw}(2019)}]{Jones2019}%
  \BibitemOpen
  \bibfield  {author} {\bibinfo {author} {\bibfnamefont {Garth~A.}\
  \bibnamefont {Jones}}\ and\ \bibinfo {author} {\bibfnamefont {David~S.}\
  \bibnamefont {Bradshaw}},\ }\bibfield  {title} {\enquote {\bibinfo {title}
  {Resonance energy transfer: From fundamental theory to recent
  applications},}\ }\href {\doibase 10.3389/fphy.2019.00100} {\bibfield
  {journal} {\bibinfo  {journal} {Front. Phys.}\ }\textbf {\bibinfo
  {volume} {7}}, \bibinfo
  {page} {100} (\bibinfo {year} {2019})}\ %10.3389/fphy.2019.00100}
  \BibitemShut
  {NoStop}%
\bibitem [{\citenamefont {Miller}(2012)}]{Miller2012}%
  \BibitemOpen
  \bibfield  {author} {\bibinfo {author} {\bibfnamefont {William~H.}\
  \bibnamefont {Miller}},\ }\bibfield  {title} {\enquote {\bibinfo {title}
  {Perspective: Quantum or classical coherence?}}\ }
  \href {\doibase 10.1063/1.4727849} {\bibfield  {journal} {\bibinfo  {journal} {J.
  Chem. Phys.}\ }\textbf {\bibinfo {volume} {136}},\ \bibinfo {pages}
  {210901} (\bibinfo {year} {2012})}\ % \Eprint
  %{http://arxiv.org/abs/https://doi.org/10.1063/1.4727849}
  %{https://doi.org/10.1063/1.4727849} 
  \BibitemShut {NoStop}%
\bibitem [{\citenamefont {{Man{\v{c}}al}}(2020)}]{Mancal2020}%
  \BibitemOpen
  \bibfield  {author} {\bibinfo {author} {\bibfnamefont {Tom{\'a}{\v{s}}}\
  \bibnamefont {{Man{\v{c}}al}}},\ }\bibfield  {title} {\enquote {\bibinfo
  {title} {{A decade with quantum coherence: How our past became classical and
  the future turned quantum}},}\ }\href {\doibase 10.1016/j.chemphys.2019.110663} {\bibfield  {journal} {\bibinfo  {journal}
  {Chem. Phys.}\ }\textbf {\bibinfo {volume} {532}},\ \bibinfo {eid}
  {110663} (\bibinfo {year} {2020})}\BibitemShut {NoStop}%
\bibitem [{\citenamefont {Mukamel}(1995)}]{Mukamel1995}%
  \BibitemOpen
  \bibfield  {author} {\bibinfo {author} {\bibfnamefont {Shaul}\ \bibnamefont
  {Mukamel}},\ }\href@noop {} {\emph {\bibinfo {title} {{Principles of
  Nonlinear Optical Spectroscopy}}}}\ (\bibinfo  {publisher} {Oxford University
  Press},\ \bibinfo {address} {New York},\ \bibinfo {year} {1995})\BibitemShut
  {NoStop}%
\bibitem [{\citenamefont {Kurt}(2021)}]{Kurt2021}%
  \BibitemOpen
  \bibfield  {author} {\bibinfo {author} {\bibfnamefont {Arzu}\ \bibnamefont
  {Kurt}},\ }\bibfield  {title} {\enquote {\bibinfo {title} {{Two-time
  correlation functions beyond quantum regression theorem: effect of external
  noise}},}\ }\href {\doibase 10.1007/s11128-021-03153-6} {\bibfield  {journal}
  {\bibinfo  {journal} {Quantum Inf. Process.}\ }\textbf {\bibinfo {volume}
  {20}},\ \bibinfo {pages} {238} (\bibinfo {year} {2021})}\ %\Eprint
  %{http://arxiv.org/abs/2101.08663} {arXiv:2101.08663} 
  \BibitemShut {NoStop}%
\bibitem [{\citenamefont {Ban}(2019)}]{Ban2019}%
  \BibitemOpen
  \bibfield  {author} {\bibinfo {author} {\bibfnamefont {Masashi}\ \bibnamefont
  {Ban}},\ }\bibfield  {title} {\enquote {\bibinfo {title} {{Two-time
  correlations functions and quantumness of an open two-level system}},}\
  }\href {\doibase 10.1140/epjd/e2018-90399-8} {\bibfield  {journal} {\bibinfo
  {journal} {Eur. Phys. J. D}\ }\textbf {\bibinfo {volume} {73}},\
  \bibinfo {eid} {12} (\bibinfo {year} {2019})}\BibitemShut {NoStop}%
\bibitem [{\citenamefont {Abo}\ \emph {et~al.}(2022)\citenamefont {Abo},
  \citenamefont {Chimczak}, \citenamefont {Kowalewska-Kud{\l}aszyk},
  \citenamefont {Pe{\v{r}}ina}, \citenamefont {Chhajlany},\ and\ \citenamefont
  {Miranowicz}}]{Abo2022}%
  \BibitemOpen
  \bibfield  {author} {\bibinfo {author} {\bibfnamefont {Shilan}\ \bibnamefont
  {Abo}}, \bibinfo {author} {\bibfnamefont {Grzegorz}\ \bibnamefont
  {Chimczak}}, \bibinfo {author} {\bibfnamefont {Anna}\ \bibnamefont
  {Kowalewska-Kud{\l}aszyk}}, \bibinfo {author} {\bibfnamefont {Jan}\
  \bibnamefont {Pe{\v{r}}ina}}, \bibinfo {author} {\bibfnamefont {Ravindra}\
  \bibnamefont {Chhajlany}}, \ and\ \bibinfo {author} {\bibfnamefont {Adam}\
  \bibnamefont {Miranowicz}},\ }\bibfield  {title} {\enquote {\bibinfo {title}
  {Hybrid photon--phonon blockade},}\ }\href {\doibase 10.1038/s41598-022-21267-4} {\bibfield  {journal} {\bibinfo  {journal}
  {Sci. Rep.}\ }\textbf {\bibinfo {volume} {12}},\ \bibinfo {pages}
  {17655} (\bibinfo {year} {2022})}\BibitemShut {NoStop}%
\bibitem [{\citenamefont {Kalaga}\ \emph {et~al.}(2022)\citenamefont {Kalaga},
  \citenamefont {Leo{\'{n}}ski}, \citenamefont {Szcz{\c{e}}{\'{s}}niak},\ and\
  \citenamefont {Pe{\v{r}}ina}}]{Kalaga2022}%
  \BibitemOpen
  \bibfield  {author} {\bibinfo {author} {\bibfnamefont {Joanna~K.}\
  \bibnamefont {Kalaga}}, \bibinfo {author} {\bibfnamefont {Wiesław}\
  \bibnamefont {Leo{\'{n}}ski}}, \bibinfo {author} {\bibfnamefont
  {Rados{\l}aw}\ \bibnamefont {Szcz{\c{e}}{\'{s}}niak}}, \ and\ \bibinfo
  {author} {\bibfnamefont {Jan}\ \bibnamefont {Pe{\v{r}}ina}},\ }\bibfield
  {title} {\enquote {\bibinfo {title} {Mixedness, coherence and entanglement in
  a family of three-qubit states},}\ }\href {\doibase 10.3390/e24030324}
  {\bibfield  {journal} {\bibinfo  {journal} {Entropy}\ }\textbf {\bibinfo
  {volume} {24}}, \bibinfo
  {page} {324} (\bibinfo {year} {2022})}\ %10.3390/e24030324}
  \BibitemShut
  {NoStop}%
\bibitem [{\citenamefont {Holzinger}\ \emph {et~al.}(2022)\citenamefont
  {Holzinger}, \citenamefont {Oh}, \citenamefont {Reitz}, \citenamefont
  {Ritsch},\ and\ \citenamefont {Genes}}]{Holzinger2022}%
  \BibitemOpen
  \bibfield  {author} {\bibinfo {author} {\bibfnamefont {R.}~\bibnamefont
  {Holzinger}}, \bibinfo {author} {\bibfnamefont {S.~A.}\ \bibnamefont {Oh}},
  \bibinfo {author} {\bibfnamefont {M.}~\bibnamefont {Reitz}}, \bibinfo
  {author} {\bibfnamefont {H.}~\bibnamefont {Ritsch}}, \ and\ \bibinfo {author}
  {\bibfnamefont {C.}~\bibnamefont {Genes}},\ }\bibfield  {title} {\enquote
  {\bibinfo {title} {Cooperative subwavelength molecular quantum emitter
  arrays},}\ }\href {\doibase 10.1103/PhysRevResearch.4.033116} {\bibfield
  {journal} {\bibinfo  {journal} {Phys. Rev. Res.}\ }\textbf {\bibinfo {volume}
  {4}},\ \bibinfo {pages} {033116} (\bibinfo {year} {2022})}\BibitemShut
  {NoStop}%
\bibitem [{\citenamefont {Pruchyathamkorn}\ \emph {et~al.}(2020)\citenamefont
  {Pruchyathamkorn}, \citenamefont {Kendrick}, \citenamefont {Frawley},
  \citenamefont {Mattioni}, \citenamefont {Caycedo-Soler}, \citenamefont
  {Huelga}, \citenamefont {Plenio},\ and\ \citenamefont
  {Anderson}}]{Pruchyathamkorn2020}%
  \BibitemOpen
  \bibfield  {author} {\bibinfo {author} {\bibfnamefont {Jiratheep}\
  \bibnamefont {Pruchyathamkorn}}, \bibinfo {author} {\bibfnamefont
  {William~J.}\ \bibnamefont {Kendrick}}, \bibinfo {author} {\bibfnamefont
  {Andrew~T.}\ \bibnamefont {Frawley}}, \bibinfo {author} {\bibfnamefont
  {Andrea}\ \bibnamefont {Mattioni}}, \bibinfo {author} {\bibfnamefont
  {Felipe}\ \bibnamefont {Caycedo-Soler}}, \bibinfo {author} {\bibfnamefont
  {Susana~F.}\ \bibnamefont {Huelga}}, \bibinfo {author} {\bibfnamefont
  {Martin~B.}\ \bibnamefont {Plenio}}, \ and\ \bibinfo {author} {\bibfnamefont
  {Harry~L.}\ \bibnamefont {Anderson}},\ }\bibfield  {title} {\enquote
  {\bibinfo {title} {A complex comprising a cyanine dye rotaxane and a
  porphyrin nanoring as a model light-harvesting system},}\ }\href {\doibase https://doi.org/10.1002/anie.202006644} {\bibfield  {journal} {\bibinfo
  {journal} {Angew. Chem. Int. Ed.}\ }\textbf {\bibinfo {volume} {59}},\
  \bibinfo {pages} {16455} (\bibinfo {year} {2020})}\ %\Eprint
  %{http://arxiv.org/abs/https://onlinelibrary.wiley.com/doi/pdf/10.1002/anie.202006644}
  %{https://onlinelibrary.wiley.com/doi/pdf/10.1002/anie.202006644} 
  \BibitemShut
  {NoStop}%
\bibitem [{\citenamefont {Breuer}\ \emph {et~al.}(2016)\citenamefont {Breuer},
  \citenamefont {Laine}, \citenamefont {Piilo},\ and\ \citenamefont
  {Vacchini}}]{Breuer2015}%
  \BibitemOpen
  \bibfield  {author} {\bibinfo {author} {\bibfnamefont {Heinz-Peter}\
  \bibnamefont {Breuer}}, \bibinfo {author} {\bibfnamefont {Elsi-Mari}\
  \bibnamefont {Laine}}, \bibinfo {author} {\bibfnamefont {Jyrki}\ \bibnamefont
  {Piilo}}, \ and\ \bibinfo {author} {\bibfnamefont {Bassano}\ \bibnamefont
  {Vacchini}},\ }\bibfield  {title} {\enquote {\bibinfo {title} {Colloquium:
  Non-markovian dynamics in open quantum systems},}\ }\href {\doibase 10.1103/RevModPhys.88.021002} {\bibfield  {journal} {\bibinfo  {journal}
  {Rev. Mod. Phys.}\ }\textbf {\bibinfo {volume} {88}},\ \bibinfo {pages}
  {021002} (\bibinfo {year} {2016})}\BibitemShut {NoStop}%
\bibitem [{\citenamefont {Breuer}\ and\ \citenamefont
  {Petruccione}(2002)}]{Breuer2002}%
  \BibitemOpen
  \bibfield  {author} {\bibinfo {author} {\bibfnamefont {Heinz-Peter~Peter}\
  \bibnamefont {Breuer}}\ and\ \bibinfo {author} {\bibfnamefont {Francesco}\
  \bibnamefont {Petruccione}},\ }\href@noop {} {\emph {\bibinfo {title} {{The
  Theory of Open Quantum Systems}}}}\ (\bibinfo  {publisher} {Oxford University
  Press},\ \bibinfo {year} {2002})\BibitemShut {NoStop}%
\bibitem [\citenamefont {Weiss}\ and\ \citenamefont
  {Weiss}(1993)]{Weiss1993}%
  \BibitemOpen
  \bibfield  {author} {\bibinfo {author} {\bibfnamefont {Ulrich}\
  \bibnamefont {Weiss}},\ }\href@noop {} {\emph {\bibinfo {title} {{Quantum Dissipative Systems}}}}\ (\bibinfo  {publisher} {Oxford University
  Press},\ \bibinfo {year} {1993})\BibitemShut {NoStop}%
\bibitem [{\citenamefont {Green}\ \emph {et~al.}(2019)\citenamefont {Green},
  \citenamefont {Humphries}, \citenamefont {Dijkstra},\ and\ \citenamefont
  {Jones}}]{Green2019}%
  \BibitemOpen
  \bibfield  {author} {\bibinfo {author} {\bibfnamefont {Dale}\ \bibnamefont
  {Green}}, \bibinfo {author} {\bibfnamefont {Ben~S.}\ \bibnamefont
  {Humphries}}, \bibinfo {author} {\bibfnamefont {Arend~G.}\ \bibnamefont
  {Dijkstra}}, \ and\ \bibinfo {author} {\bibfnamefont {Garth~A.}\ \bibnamefont
  {Jones}},\ }\bibfield  {title} {\enquote {\bibinfo {title} {Quantifying
  non-markovianity in underdamped versus overdamped environments and its effect
  on spectral lineshape},}\ }\href {\doibase 10.1063/1.5119300} {\bibfield
  {journal} {\bibinfo  {journal} {J. Chem. Phys.}\ }\textbf
  {\bibinfo {volume} {151}},\ \bibinfo {pages} {174112} (\bibinfo {year}
  {2019})}\ %\Eprint {http://arxiv.org/abs/https://doi.org/10.1063/1.5119300}
  %{https://doi.org/10.1063/1.5119300} 
  \BibitemShut {NoStop}%
\bibitem [{\citenamefont {Humphries}\ \emph {et~al.}(2022)\citenamefont
  {Humphries}, \citenamefont {Green},\ and\ \citenamefont
  {Jones}}]{Humphries2022}%
  \BibitemOpen
  \bibfield  {author} {\bibinfo {author} {\bibfnamefont {Ben~S.}\ \bibnamefont
  {Humphries}}, \bibinfo {author} {\bibfnamefont {Dale}\ \bibnamefont {Green}},
  \ and\ \bibinfo {author} {\bibfnamefont {Garth~A.}\ \bibnamefont {Jones}},\
  }\bibfield  {title} {\enquote {\bibinfo {title} {The influence of a
  hamiltonian vibration vs a bath vibration on the 2d electronic spectra of a
  homodimer},}\ }\href {\doibase 10.1063/5.0077404} {\bibfield  {journal}
  {\bibinfo  {journal} {J. Chem. Phys.}\ }\textbf {\bibinfo
  {volume} {156}}\ \bibinfo {pages} {084103} (\bibinfo {year} {2022})},\
  %\Eprint {http://arxiv.org/abs/https://doi.org/10.1063/5.0077404}
  %{https://doi.org/10.1063/5.0077404} 
  \BibitemShut {NoStop}%
\bibitem [{\citenamefont {Mollow}(1969)}]{Mollow1969}%
  \BibitemOpen
  \bibfield  {author} {\bibinfo {author} {\bibfnamefont {B.~R.}\ \bibnamefont
  {Mollow}},\ }\bibfield  {title} {\enquote {\bibinfo {title} {Power spectrum
  of light scattered by two-level systems},}\ }\href {\doibase 10.1103/PhysRev.188.1969} {\bibfield  {journal} {\bibinfo  {journal} {Phys.
  Rev.}\ }\textbf {\bibinfo {volume} {188}},\ \bibinfo {pages} {1969}
  (\bibinfo {year} {1969})}\BibitemShut {NoStop}%
\bibitem [{\citenamefont {Glauber}(1963)}]{Glauber1963}%
  \BibitemOpen
  \bibfield  {author} {\bibinfo {author} {\bibfnamefont {Roy~J.}\ \bibnamefont
  {Glauber}},\ }\bibfield  {title} {\enquote {\bibinfo {title} {The quantum
  theory of optical coherence},}\ }\href {\doibase 10.1103/PhysRev.130.2529}
  {\bibfield  {journal} {\bibinfo  {journal} {Phys. Rev.}\ }\textbf {\bibinfo
  {volume} {130}},\ \bibinfo {pages} {2529} (\bibinfo {year}
  {1963})}\BibitemShut {NoStop}%
\bibitem [{\citenamefont {Egorova}\ \emph {et~al.}(2007)\citenamefont
  {Egorova}, \citenamefont {Gelin},\ and\ \citenamefont
  {Domcke}}]{Egorova2007}%
  \BibitemOpen
  \bibfield  {author} {\bibinfo {author} {\bibfnamefont {Dassia}\ \bibnamefont
  {Egorova}}, \bibinfo {author} {\bibfnamefont {Maxim~F.}\ \bibnamefont
  {Gelin}}, \ and\ \bibinfo {author} {\bibfnamefont {Wolfgang}\ \bibnamefont
  {Domcke}},\ }\bibfield  {title} {\enquote {\bibinfo {title} {Analysis of
  cross peaks in two-dimensional electronic photon-echo spectroscopy for simple
  models with vibrations and dissipation},}\ }\href {\doibase 10.1063/1.2435353} {\bibfield  {journal} {\bibinfo  {journal} {J.
  Chem. Phys.}\ }\textbf {\bibinfo {volume} {126}},\ \bibinfo {pages}
  {074314} (\bibinfo {year} {2007})}\ % \Eprint
  %{http://arxiv.org/abs/https://doi.org/10.1063/1.2435353}
  %{https://doi.org/10.1063/1.2435353} 
  \BibitemShut {NoStop}%
\bibitem [{sup()}]{supplement}%
  \BibitemOpen
  \href@noop {} {}\bibinfo {note} See Supplemental Material at [URL will be
  inserted by publisher] for non-normalised correlation functions and
  discussion of bases, which includes Ref. \cite{Andrews1996}.\BibitemShut {Stop}%
\bibitem [{\citenamefont {Andrews}\ \emph {et~al.}(1996)\citenamefont
  {Andrews}, \citenamefont {Mewes}, \citenamefont {van Druten}, \citenamefont
  {Durfee}, \citenamefont {Kurn},\ and\ \citenamefont
  {Ketterle}}]{Andrews1996}%
  \BibitemOpen
  \bibfield  {author} {\bibinfo {author} {\bibfnamefont {M.~R.}\ \bibnamefont
  {Andrews}}, \bibinfo {author} {\bibfnamefont {M.-O.}\ \bibnamefont {Mewes}},
  \bibinfo {author} {\bibfnamefont {N.~J.}\ \bibnamefont {van Druten}},
  \bibinfo {author} {\bibfnamefont {D.~S.}\ \bibnamefont {Durfee}}, \bibinfo
  {author} {\bibfnamefont {D.~M.}\ \bibnamefont {Kurn}},\ and\ \bibinfo
  {author} {\bibfnamefont {W.}~\bibnamefont {Ketterle}},\ }\bibfield  {title}
  {\bibinfo {title} {\enquote{Direct, Nondestructive Observation of a Bose
  Condensate}},\ }\href {https://doi.org/10.1126/science.273.5271.84}
  {\bibfield  {journal} {\bibinfo  {journal} {Science}\ }\textbf {\bibinfo
  {volume} {273}},\ \bibinfo {pages} {84} (\bibinfo {year} {1996})}\BibitemShut
  {NoStop}%
\bibitem [{\citenamefont {Hamm}\ and\ \citenamefont {Zanni}(2011)}]{Hamm2011}%
  \BibitemOpen
  \bibfield  {author} {\bibinfo {author} {\bibfnamefont {Peter}\ \bibnamefont
  {Hamm}}\ and\ \bibinfo {author} {\bibfnamefont {Martin}\ \bibnamefont
  {Zanni}},\ }\href {\doibase 10.1017/CBO9780511675935} {\emph {\bibinfo
  {title} {{Concepts and Methods of 2D Infrared Spectroscopy}}}}\ (\bibinfo
  {publisher} {Cambridge University Press},\ \bibinfo {year}
  {2011})\BibitemShut {NoStop}%
\bibitem [{\citenamefont {{Ishizaki}}\ and\ \citenamefont
  {{Tanimura}}(2008)}]{Ishizaki2008}%
  \BibitemOpen
  \bibfield  {author} {\bibinfo {author} {\bibfnamefont {Akihito}\ \bibnamefont
  {{Ishizaki}}}\ and\ \bibinfo {author} {\bibfnamefont {Yoshitaka}\
  \bibnamefont {{Tanimura}}},\ }\bibfield  {title} {\enquote {\bibinfo {title}
  {{Nonperturbative non-Markovian quantum master equation: Validity and
  limitation to calculate nonlinear response functions}},}\ }\href {\doibase 10.1016/j.chemphys.2007.10.037} {\bibfield  {journal} {\bibinfo  {journal}
  {Chem. Phys.}\ }\textbf {\bibinfo {volume} {347}},\ \bibinfo {pages}
  {185} (\bibinfo {year} {2008})}\BibitemShut {NoStop}%
\bibitem [{\citenamefont {Carballeira}\ \emph {et~al.}(2021)\citenamefont
  {Carballeira}, \citenamefont {Dolgitzer}, \citenamefont {Zhao}, \citenamefont
  {Zeng},\ and\ \citenamefont {Chen}}]{Carballeira2021}%
  \BibitemOpen
  \bibfield  {author} {\bibinfo {author} {\bibfnamefont {Rafael}\ \bibnamefont
  {Carballeira}}, \bibinfo {author} {\bibfnamefont {David}\ \bibnamefont
  {Dolgitzer}}, \bibinfo {author} {\bibfnamefont {Peng}\ \bibnamefont {Zhao}},
  \bibinfo {author} {\bibfnamefont {Debing}\ \bibnamefont {Zeng}}, \ and\
  \bibinfo {author} {\bibfnamefont {Yusui}\ \bibnamefont {Chen}},\ }\bibfield
  {title} {\enquote {\bibinfo {title} {Stochastic schr{\"o}dinger equation
  derivation of non-markovian two-time correlation functions},}\ }\href
  {\doibase 10.1038/s41598-021-91216-0} {\bibfield  {journal} {\bibinfo
  {journal} {Sci. Rep.}\ }\textbf {\bibinfo {volume} {11}},\ \bibinfo
  {pages} {11828} (\bibinfo {year} {2021})}\BibitemShut {NoStop}%
\bibitem [{\citenamefont {{Alkathiri}}\ \emph {et~al.}(2022)\citenamefont
  {{Alkathiri}}, \citenamefont {{Alsallami}}, \citenamefont {{Abdel-Wahab}},
  \citenamefont {{Abdel-Khalek}},\ and\ \citenamefont
  {{Khalil}}}]{Alkathiri2022}%
  \BibitemOpen
  \bibfield  {author} {\bibinfo {author} {\bibfnamefont {Ali~A.}\ \bibnamefont
  {{Alkathiri}}}, \bibinfo {author} {\bibfnamefont {Shami A.~M.}\ \bibnamefont
  {{Alsallami}}}, \bibinfo {author} {\bibfnamefont {N.~H.}\ \bibnamefont
  {{Abdel-Wahab}}}, \bibinfo {author} {\bibfnamefont {S.}~\bibnamefont
  {{Abdel-Khalek}}}, \ and\ \bibinfo {author} {\bibfnamefont {E.~M.}\
  \bibnamefont {{Khalil}}},\ }\bibfield  {title} {\enquote {\bibinfo {title}
  {{On the interaction between {\ensuremath{\Lambda}}-type five-level atom and
  one-mode squeezed coherent field}},}\ }\href {\doibase 10.1016/j.rinp.2022.105739} {\bibfield  {journal} {\bibinfo  {journal}
  {Results Phys.}\ }\textbf {\bibinfo {volume} {39}},\ \bibinfo {eid}
  {105739} (\bibinfo {year} {2022})}\BibitemShut {NoStop}%
\bibitem [{\citenamefont {Yang}\ \emph {et~al.}(2022)\citenamefont {Yang},
  \citenamefont {Čufar}, \citenamefont {Pahl},\ and\ \citenamefont
  {Brand}}]{Yang2022}%
  \BibitemOpen
  \bibfield  {author} {\bibinfo {author} {\bibfnamefont {Mingrui}\ \bibnamefont
  {Yang}}, \bibinfo {author} {\bibfnamefont {Matija}\ \bibnamefont {Čufar}},
  \bibinfo {author} {\bibfnamefont {Elke}\ \bibnamefont {Pahl}}, \ and\
  \bibinfo {author} {\bibfnamefont {Joachim}\ \bibnamefont {Brand}},\
  }\bibfield  {title} {\enquote {\bibinfo {title} {Polaron-depleton transition
  in the yrast excitations of a one-dimensional bose gas with a mobile
  impurity},}\ }\href {\doibase 10.3390/condmat7010015} {\bibfield  {journal}
  {\bibinfo  {journal} {Condens. Matter}\ }\textbf {\bibinfo {volume} {7}}
  (\bibinfo {year} {2022})}%\ 10.3390/condmat7010015}
  \BibitemShut {NoStop}%
\bibitem [{\citenamefont {Tanimura}\ and\ \citenamefont
  {Kubo}(1989)}]{Tanimura1989}%
  \BibitemOpen
  \bibfield  {author} {\bibinfo {author} {\bibfnamefont {Yoshitaka}\
  \bibnamefont {Tanimura}}\ and\ \bibinfo {author} {\bibfnamefont {Ryogo}\
  \bibnamefont {Kubo}},\ }\bibfield  {title} {\enquote {\bibinfo {title} {Time
  evolution of a quantum system in contact with a nearly gaussian-markoffian
  noise bath},}\ }\href {\doibase 10.1143/JPSJ.58.101} {\bibfield  {journal}
  {\bibinfo  {journal} {J. Phys. Soc. Jpn.}\ }\textbf
  {\bibinfo {volume} {58}},\ \bibinfo {pages} {101} (\bibinfo {year}
  {1989})}\ % \Eprint {http://arxiv.org/abs/https://doi.org/10.1143/JPSJ.58.101}
  %{https://doi.org/10.1143/JPSJ.58.101} 
  \BibitemShut {NoStop}%
\bibitem [{\citenamefont {Tanimura}(2020)}]{Tanimura2020}%
  \BibitemOpen
  \bibfield  {author} {\bibinfo {author} {\bibfnamefont {Yoshitaka}\
  \bibnamefont {Tanimura}},\ }\bibfield  {title} {\enquote {\bibinfo {title}
  {Numerically “exact” approach to open quantum dynamics: The hierarchical
  equations of motion (heom)},}\ }\href {\doibase 10.1063/5.0011599} {\bibfield
   {journal} {\bibinfo  {journal} {J Chem. Phys.}\ }\textbf
  {\bibinfo {volume} {153}},\ \bibinfo {pages} {020901} (\bibinfo {year}
  {2020})}\ %\Eprint {http://arxiv.org/abs/https://doi.org/10.1063/5.0011599}
  %{https://doi.org/10.1063/5.0011599} 
  \BibitemShut {NoStop}%
\bibitem [{\citenamefont {Fano}(1961)}]{Fano1961}%
  \BibitemOpen
  \bibfield  {author} {\bibinfo {author} {\bibfnamefont {U.}~\bibnamefont
  {Fano}},\ }\bibfield  {title} {\enquote {\bibinfo {title} {Quantum theory of
  interference effects in the mixing of light from phase-independent
  sources},}\ }\href {\doibase 10.1119/1.1937827} {\bibfield  {journal}
  {\bibinfo  {journal} {Am. J. Phys.}\ }\textbf {\bibinfo
  {volume} {29}},\ \bibinfo {pages} {539} (\bibinfo {year} {1961})}\
  %\Eprint {http://arxiv.org/abs/https://doi.org/10.1119/1.1937827}
  %{https://doi.org/10.1119/1.1937827} 
  \BibitemShut {NoStop}%
\bibitem [{\citenamefont {Lax}(1963)}]{Lax1963}%
  \BibitemOpen
  \bibfield  {author} {\bibinfo {author} {\bibfnamefont {Melvin}\ \bibnamefont
  {Lax}},\ }\bibfield  {title} {\enquote {\bibinfo {title} {Formal theory of
  quantum fluctuations from a driven state},}\ }\href {\doibase 10.1103/PhysRev.129.2342} {\bibfield  {journal} {\bibinfo  {journal} {Phys.
  Rev.}\ }\textbf {\bibinfo {volume} {129}},\ \bibinfo {pages} {2342}
  (\bibinfo {year} {1963})}\BibitemShut {NoStop}%
\bibitem [{\citenamefont {Lax}(1967)}]{Lax1967}%
  \BibitemOpen
  \bibfield  {author} {\bibinfo {author} {\bibfnamefont {Melvin}\ \bibnamefont
  {Lax}},\ }\bibfield  {title} {\enquote {\bibinfo {title} {Quantum noise. x.
  density-matrix treatment of field and population-difference fluctuations},}\
  }\href {\doibase 10.1103/PhysRev.157.213} {\bibfield  {journal} {\bibinfo
  {journal} {Phys. Rev.}\ }\textbf {\bibinfo {volume} {157}},\ \bibinfo {pages}
  {213} (\bibinfo {year} {1967})}\BibitemShut {NoStop}%
\bibitem [{\citenamefont {Carmichael}(1999)}]{Carmichael1999}%
  \BibitemOpen
  \bibfield  {author} {\bibinfo {author} {\bibfnamefont {Howard~J.}\
  \bibnamefont {Carmichael}},\ }\href {\doibase 10.1007/978-3-662-03875-8}
  {\emph {\bibinfo {title} {Statistical Methods in Quantum Optics 1}}}\
  (\bibinfo  {publisher} {Springer Berlin Heidelberg},\ \bibinfo {year}
  {1999})\ pp.\ \bibinfo {pages} {19}\BibitemShut {NoStop}%
\bibitem [{\citenamefont {Camargo}\ \emph {et~al.}(2015)\citenamefont
  {Camargo}, \citenamefont {Anderson}, \citenamefont {Meech},\ and\
  \citenamefont {Heisler}}]{Camargo2015}%
  \BibitemOpen
  \bibfield  {author} {\bibinfo {author} {\bibfnamefont {F.~V.}\ \bibnamefont
  {Camargo}}, \bibinfo {author} {\bibfnamefont {H.~L.}\ \bibnamefont
  {Anderson}}, \bibinfo {author} {\bibfnamefont {S.~R.}\ \bibnamefont
  {Meech}},\ and\ \bibinfo {author} {\bibfnamefont {I.~A.}\ \bibnamefont
  {Heisler}},\ }\bibfield  {title} {\bibinfo {title} {{Full characterization of
  vibrational coherence in a porphyrin chromophore by two-dimensional
  electronic spectroscopy}},\ }\href {https://doi.org/10.1021/jp511881a}
  {\bibfield  {journal} {\bibinfo  {journal} {J. Phys. Chem. A}\ }\textbf
  {\bibinfo {volume} {119}},\ \bibinfo {pages} {95} (\bibinfo {year}
  {2015})}\BibitemShut {NoStop}%
\bibitem [{\citenamefont {Lu}\ \emph {et~al.}(2020)\citenamefont {Lu},
  \citenamefont {Lee},\ and\ \citenamefont {Anna}}]{Lu2020}%
  \BibitemOpen
  \bibfield  {author} {\bibinfo {author} {\bibfnamefont {J.}~\bibnamefont
  {Lu}}, \bibinfo {author} {\bibfnamefont {Y.}~\bibnamefont {Lee}},\ and\
  \bibinfo {author} {\bibfnamefont {J.~M.}\ \bibnamefont {Anna}},\ }\bibfield
  {title} {\bibinfo {title} {{Extracting the Frequency-Dependent Dynamic Stokes
  Shift from Two-Dimensional Electronic Spectra with Prominent Vibrational
  Coherences}},\ }\href {https://doi.org/10.1021/acs.jpcb.0c05522} {\bibfield
  {journal} {\bibinfo  {journal} {J. Phys. Chem. B}\ }\textbf {\bibinfo
  {volume} {124}},\ \bibinfo {pages} {8857} (\bibinfo {year}
  {2020})}\BibitemShut {NoStop}%
\bibitem [{\citenamefont {Nafie}(1983)}]{Nafie1983}%
  \BibitemOpen
  \bibfield  {author} {\bibinfo {author} {\bibfnamefont {Laurence~A.}\
  \bibnamefont {Nafie}},\ }\bibfield  {title} {\enquote {\bibinfo {title}
  {Adiabatic molecular properties beyond the born–oppenheimer approximation.
  complete adiabatic wave functions and vibrationally induced electronic
  current density},}\ }\href {\doibase 10.1063/1.445588} {\bibfield  {journal}
  {\bibinfo  {journal} {J. Chem. Phys.}\ }\textbf {\bibinfo
  {volume} {79}},\ \bibinfo {pages} {4950} (\bibinfo {year} {1983})}\
  %\Eprint {http://arxiv.org/abs/https://doi.org/10.1063/1.445588}
  %{https://doi.org/10.1063/1.445588} 
  \BibitemShut {NoStop}%
\bibitem [{\citenamefont {Mal{\'{y}}}\ \emph {et~al.}(2016)\citenamefont
  {Mal{\'{y}}}, \citenamefont {Somsen}, \citenamefont {Novoderezhkin},
  \citenamefont {Man{\v{c}}al},\ and\ \citenamefont {{Van
  Grondelle}}}]{Maly2016}%
  \BibitemOpen
  \bibfield  {author} {\bibinfo {author} {\bibfnamefont {Pavel}\ \bibnamefont
  {Mal{\'{y}}}}, \bibinfo {author} {\bibfnamefont {Oscar~J.G.}\ \bibnamefont
  {Somsen}}, \bibinfo {author} {\bibfnamefont {Vladimir~I.}\ \bibnamefont
  {Novoderezhkin}}, \bibinfo {author} {\bibfnamefont {Tom{\'{a}}{\v{s}}}\
  \bibnamefont {Man{\v{c}}al}}, \ and\ \bibinfo {author} {\bibfnamefont
  {Rienk}\ \bibnamefont {{Van Grondelle}}},\ }\bibfield  {title} {\enquote
  {\bibinfo {title} {The role of resonant vibrations in electronic energy
  transfer},}\ }\href {\doibase https://doi.org/10.1002/cphc.201500965}
  {\bibfield  {journal} {\bibinfo  {journal} {ChemPhysChem}\ }\textbf {\bibinfo
  {volume} {17}},\ \bibinfo {pages} {1356} (\bibinfo {year}
  {2016})}\BibitemShut {NoStop}%
\bibitem [{\citenamefont {Kopec}\ \emph {et~al.}(2012)\citenamefont {Kopec},
  \citenamefont {Ottiger}, \citenamefont {Leutwyler},\ and\ \citenamefont
  {K{\"{o}}ppel}}]{Kopec2012}%
  \BibitemOpen
  \bibfield  {author} {\bibinfo {author} {\bibfnamefont {Sabine}\ \bibnamefont
  {Kopec}}, \bibinfo {author} {\bibfnamefont {Philipp}\ \bibnamefont
  {Ottiger}}, \bibinfo {author} {\bibfnamefont {Samuel}\ \bibnamefont
  {Leutwyler}}, \ and\ \bibinfo {author} {\bibfnamefont {Horst}\ \bibnamefont
  {K{\"{o}}ppel}},\ }\bibfield  {title} {\enquote {\bibinfo {title}
  {Vibrational quenching of excitonic splittings in h-bonded molecular dimers:
  Adiabatic description and effective mode approximation},}\ }\href {\doibase 10.1063/1.4763979} {\bibfield  {journal} {\bibinfo  {journal} {J.
  Chem. Phys.}\ }\textbf {\bibinfo {volume} {137}},\ \bibinfo {pages}
  {184312} (\bibinfo {year} {2012})}\ %\Eprint
  %{http://arxiv.org/abs/https://doi.org/10.1063/1.4763979}
  %{https://doi.org/10.1063/1.4763979} 
  \BibitemShut {NoStop}%
\bibitem [{\citenamefont {Condon}(1926)}]{Condon1926}%
  \BibitemOpen
  \bibfield  {author} {\bibinfo {author} {\bibfnamefont {Edward}\ \bibnamefont
  {Condon}},\ }\bibfield  {title} {\enquote {\bibinfo {title} {A theory of
  intensity distribution in band systems},}\ }\href {\doibase 10.1103/PhysRev.28.1182} {\bibfield  {journal} {\bibinfo  {journal} {Phys.
  Rev.}\ }\textbf {\bibinfo {volume} {28}},\ \bibinfo {pages} {1182}
  (\bibinfo {year} {1926})}\BibitemShut {NoStop}%
\bibitem [{\citenamefont {Rabi}(1937)}]{Rabi1937}%
  \BibitemOpen
  \bibfield  {author} {\bibinfo {author} {\bibfnamefont {I.~I.}\ \bibnamefont
  {Rabi}},\ }\bibfield  {title} {\enquote {\bibinfo {title} {Space quantization
  in a gyrating magnetic field},}\ }\href {\doibase 10.1103/PhysRev.51.652}
  {\bibfield  {journal} {\bibinfo  {journal} {Phys. Rev.}\ }\textbf {\bibinfo
  {volume} {51}},\ \bibinfo {pages} {652} (\bibinfo {year}
  {1937})}\BibitemShut {NoStop}%
\bibitem [{\citenamefont {Mandel}\ and\ \citenamefont
  {Wolf}(1995)}]{Mandel1995}%
  \BibitemOpen
  \bibfield  {author} {\bibinfo {author} {\bibfnamefont {Leonard}\ \bibnamefont
  {Mandel}}\ and\ \bibinfo {author} {\bibfnamefont {Emil}\ \bibnamefont
  {Wolf}},\ }\href {\doibase 10.1017/CBO9781139644105} {\emph {\bibinfo {title}
  {{Optical Coherence and Quantum Optics}}}}\ (\bibinfo  {publisher} {Cambridge
  University Press},\ \bibinfo {year} {1995})\BibitemShut {NoStop}%
\bibitem [{\citenamefont {Scully}\ and\ \citenamefont
  {Zubairy}(1997)}]{Scully1997}%
  \BibitemOpen
  \bibfield  {author} {\bibinfo {author} {\bibfnamefont {Marlan~O.}\
  \bibnamefont {Scully}}\ and\ \bibinfo {author} {\bibfnamefont {M.~Suhail}\
  \bibnamefont {Zubairy}},\ }\href {\doibase 10.1017/CBO9780511813993} {\emph
  {\bibinfo {title} {Quantum Optics}}}\ (\bibinfo  {publisher} {Cambridge
  University Press},\ \bibinfo {year} {1997})\BibitemShut {NoStop}%
%
%
%
\bibitem [{\citenamefont {Brash}\ \emph {et~al.}(2019)\citenamefont {Brash},
  \citenamefont {Iles-Smith}, \citenamefont {Phillips}, \citenamefont
  {McCutcheon}, \citenamefont {O'Hara}, \citenamefont {Clarke}, \citenamefont
  {Royall}, \citenamefont {Wilson}, \citenamefont {M{\o}rk}, \citenamefont
  {Skolnick}, \citenamefont {Fox},\ and\ \citenamefont {Nazir}}]{Brash2019}%
  \BibitemOpen
  \bibfield  {author} {\bibinfo {author} {\bibfnamefont {Alistair~J.}\
  \bibnamefont {Brash}}, \bibinfo {author} {\bibfnamefont {Jake}\ \bibnamefont
  {Iles-Smith}}, \bibinfo {author} {\bibfnamefont {Catherine~L.}\ \bibnamefont
  {Phillips}}, \bibinfo {author} {\bibfnamefont {Dara~P.S.}\ \bibnamefont
  {McCutcheon}}, \bibinfo {author} {\bibfnamefont {John}\ \bibnamefont
  {O'Hara}}, \bibinfo {author} {\bibfnamefont {Edmund}\ \bibnamefont {Clarke}},
  \bibinfo {author} {\bibfnamefont {Benjamin}\ \bibnamefont {Royall}}, \bibinfo
  {author} {\bibfnamefont {Luke~R.}\ \bibnamefont {Wilson}}, \bibinfo {author}
  {\bibfnamefont {Jesper}\ \bibnamefont {M{\o}rk}}, \bibinfo {author}
  {\bibfnamefont {Maurice~S.}\ \bibnamefont {Skolnick}}, \bibinfo {author}
  {\bibfnamefont {A.~Mark}\ \bibnamefont {Fox}}, \ and\ \bibinfo {author}
  {\bibfnamefont {Ahsan}\ \bibnamefont {Nazir}},\ }\bibfield  {title} {\enquote
  {\bibinfo {title} {{Light Scattering from Solid-State Quantum Emitters:
  Beyond the Atomic Picture}},}\ }\href {\doibase 10.1103/PhysRevLett.123.167403} {\bibfield  {journal} {\bibinfo  {journal}
  {Phys. Rev. Lett.}\ }\textbf {\bibinfo {volume} {123}},\ \bibinfo{pages}{167403} (\bibinfo {year}
  {2019})
  %,\ 10.1103/PhysRevLett.123.167403
  }\BibitemShut {NoStop}%
\bibitem [{\citenamefont {Humphries}\ \emph {et~al.}(2023)\citenamefont {Humphries}, \citenamefont {Green}, \citenamefont {Borgh},\ and\ \citenamefont{Jones}}]       {Humphries2023}% 
\BibitemOpen
\bibfield {author} {\bibinfo {author} {\bibfnamefont {Ben~S.}\ \bibnamefont{Humphries}}, \bibinfo{author} {\bibfnamefont {Dale}\ \bibnamefont{Green}}, \bibinfo{author} {\bibfnamefont {Magnus~O.}\ \bibnamefont{Borgh}}, \ and\ \bibinfo{author} {\bibfnamefont {Garth~A.}\ \bibnamefont{Jones}},\ }\bibfield {title} {\bibinfo {title} {{Datasets from \enquote{Phonon Signatures in Photon Correlations} (2023)},}\ }\href {https://doi.org/10.5281/zenodo.8325538} {\bibfield {database} {\bibinfo {database} {10.5281/zenodo.8325538}}}
\BibitemShut {NoStop}%
\end{thebibliography}

%

\end{document}

%New commands
%%% custom supplemental style %%
\renewcommand{\thesection}{S\Roman{section}}
\renewcommand{\thesubsection}{S\Roman{section}.\Alph{subsection}}
\renewcommand{\theequation}{S\arabic{equation}}
\renewcommand{\thefigure}{S\arabic{figure}}
\renewcommand{\bibnumfmt}[1]{[S#1]}
\renewcommand{\citenumfont}[1]{S#1}

\title{Supplemental Material to ``Phonon signatures in photon correlations."}

\author{Ben S. Humphries}
\affiliation{School of Chemistry, University of East Anglia, Norwich Research Park, Norwich, NR4 7TJ, United Kingdom}
\author{Dale Green}
\affiliation{School of Chemistry, University of East Anglia, Norwich Research Park, Norwich, NR4 7TJ, United Kingdom}
\author{Magnus O. Borgh}
\email{M.Borgh@uea.ac.uk}
\affiliation{Physics, Faculty of Science, University of East Anglia, Norwich NR4 7TJ, United Kingdom}
\author{Garth A. Jones}
\email{garth.jones@uea.ac.uk}
\affiliation{School of Chemistry, University of East Anglia, Norwich Research Park, Norwich, NR4 7TJ, United Kingdom}

\keywords{Markovianity, HEOM, G2 two-time correlation function, phonon G2 }
\maketitle

\section{Diabatic and Adiabatic bases}
We start with the vibrational Hamiltonian,
    \begin{equation}
        \hat{H}_{\mathrm{S}}= |g\rangle \hat{h}_{g} \langle g|+|e\rangle \hat{h}_{e}\langle e|. \label{MukHs}
    \end{equation}
\begin{equation}
    \hat{h}_{g} = \frac{\hat{p}^2}{2m} + \frac{m\omega^2 \hat{q}^2}{2},
\end{equation}
\begin{equation}
    \hat{h}_{e} = \hbar\omega_{eg} + \frac{\hat{p}^2}{2m} + \frac{m\omega^2 }{2}(\hat{q}-d)^2,
\end{equation}
where $m$, $\hat{p}$, $\hat{q}$ are the mass, momentum, and coordinate, respectively, of the vibrational mode with frequency $\omega$, and $d$ is the displacement of the excited state potential. Measuring lengths in units of the oscillator length, we find the dimensionless momentum and coordinate
\begin{equation}
    \hat{P} = \frac{\hat{p}}{\sqrt{\hbar m \omega}}, \qquad  \hat{Q} = \sqrt{\frac{m\omega}{\hbar}}\hat{q},
\end{equation}
and correspondingly define
\begin{equation}
    \Delta = \sqrt{\frac{m\omega}{\hbar}}d, \qquad     \hat{b} = \frac{\big{[}\hat{Q}+i\hat{P}\big{]}}{\sqrt{2}},
\end{equation}
from which the nuclear Hamiltonians for the ground and excited states take the form
\begin{align}
    \label{suppleq:hg}
    \hat{h}_{g} &= \hbar\omega_0\left(\hat{b}^{\dagger} \hat{b}+\tfrac{1}{2}\right), \\
    \label{suppleq:he}
    \hat{h}_{e} &=  \hbar(\omega_{eg}+\lambda)+\hbar\omega_0\left[\hat{b}^{\dagger} \hat{b}-\frac{\Delta}{\sqrt{2}}(\hat{b}+\hat{b}^{\dagger})+\tfrac{1}{2}\right].
\end{align} 
The ground-state nuclear Hamiltonian~\eqref{suppleq:hg} is simply a harmonic oscillator whose eigenstates $\ket{n}$ ($n=0,1,\ldots$) form an orthonormal basis of vibrational levels. For practical computations, the Hilbert space is truncated, $n=0,1,\ldots,N$, with $N$ chosen such that the results are insensitive to the truncation.  The ground state is by definition diagonal in this basis, but the excited state acquires off-diagonal coupling terms proportional to $(\hat{b}+\hat{b}^{\dagger})$. Combining with the electronic degrees of freedom yields the full set of basis states in the diabatic basis:
\begin{equation}
    \ket{\psi^{\mathrm{D}}} = \ket{\alpha}\otimes\ket{n} = \ket{\alpha,n},
\end{equation}
where $\alpha=g,e$. The total wavefunction is then the linear combination
\begin{equation}
    \ket{\Psi} = \sum_k c_k\ket{\psi^{\mathrm{D}}_k},
\end{equation}
with complex coefficients $c_k$.  Note that in the diabatic basis, the only  coupling between electronic and vibrational motion arises from the off-diagonal terms in Eq.~\eqref{suppleq:he}.

The diabatic basis provides an intuitive physical picture. However, experimentally relevant observables correspond to an adiabatic (molecular) basis: 
For any excited-state displacement $\Delta>0$, the Hamiltonian is diagonalized by a unitary transformation
\begin{equation}
    \hat{U}^{\mathrm{AD}} = \sum_k \ket{\psi^{\mathrm{A}}_k}\bra{\psi^{\mathrm{D}}_k},
\end{equation}
such that
\begin{equation}
    \hat{H}^{\mathrm{D}}_{\mathrm{tot}}=\sum_k\epsilon_k\ket{\psi_k^{\mathrm{A}}}\bra{\psi_k^{\mathrm{A}}},
\end{equation}
where the eigenstates $\ket{\psi^{\mathrm{A}}_k}$, with eigenvalues $\epsilon_k$, form the adiabatic basis states. The adiabatic Hamiltonian is then constructed as
\begin{equation}\label{U_rotation}
    \hat{H}^{\mathrm{A}}_{\mathrm{tot}} = (\hat{U}^{\mathrm{AD}})^{\dagger}\hat{H}_{\mathrm{tot}}^{\mathrm{D}}\hat{U}^{\mathrm{AD}} = \sum_k \epsilon_k \ket{\psi^{\mathrm{D}}_k}\bra{\psi^{\mathrm{D}}_k}.
\end{equation}
On matrix form, the diabatic Hamiltonian
\begin{equation}
    \hat{H}^{\mathrm{D}}_{\mathrm{tot}} =
\begin{pmatrix}
     \hat{h}_{g} & \hat{0} \\
     \hat{0} & \hat{h}_{e}
\end{pmatrix},
\end{equation}
is thus replaced by the adiabatic
\begin{equation}\label{adiab_H}
    \hat{H}^{\mathrm{A}}_{\mathrm{tot}} =
\begin{pmatrix}
     \hat{h}_{g} & \hat{0} \\
     \hat{0} & \hat{B}
\end{pmatrix},
\end{equation}
where
\begin{equation}
    \hat{B} = \hbar(\omega_{eg}+\lambda)+\hbar\omega_0\left(\hat{b}^{\dagger} \hat{b}+\tfrac{1}{2}\right).
\end{equation}
Note that the energies of the adiabatic basis states coincide with the diabatic ones, and that  
$\hat{B}$ is in fact $\hat{h}_{e}$ without the off-diagonal elements. 

\section{Initialization to a Boltzmann Distribution}
The population in our system is initialised to the thermal equilibrium distribution over vibrational modes in the electronic ground state before the calculations are performed. The initial density matrix for the monomer is then
\begin{equation}\label{dens_a}
    \hat{\rho}^{\mathrm{A}} = \sum_k P_k\ket{\psi^{\mathrm{D}}_{k}}\bra{\psi^{\mathrm{D}}_{k}},
\end{equation}
where the weights
\begin{equation}
    P_k = \frac{\exp(-\frac{\epsilon_k}{k_{B}T})}{Z}
\end{equation}
are given by the Boltzmann distribution with  canonical partition function
\begin{equation}
    Z = \sum_k\exp\left(-\frac{\epsilon_k}{k_B T}\right) = \mathrm{Tr} \left[ \exp\left(-\frac{\hat{H}_{\mathrm{tot}}^{\mathrm{A}}}{k_B T}\right)\right].
\end{equation}
Here $\epsilon_k$ is the energy of the $k$th vibrational level in the electronic ground state. The diabatic density matrix is then calculated by rotating equation \ref{dens_a}
\begin{equation}
    \hat{\rho}^{\mathrm{D}}=\hat{U}^{\mathrm{AD}}\hat{\rho}^{\mathrm{A}}\hat({U}^{\mathrm{AD}})^{\dagger}.
\end{equation}
Calculations are then run in the diabatic basis with a Hilbert space truncation of $N=10$.

\section{Simultaneous time correlation functions and non-normalised equivalents}

\noindent In order to calculate correlation functions we have assumed that the incident field is not detected as part of the scattered field. Experimentally can be achieved, for example, by addition of a thin wire as in the dark-ground imaging technique of reference (\citenum{Andrews1996}).

The detection probability for photons is given by, 
\begin{equation}
    g_a^{(1)}(t,\tau)=\frac{\mathrm{Tr}\left[\hat{a}\exp(\mathcal{L}\tau)(\hat{\rho}) \hat{a}^{\dagger}\right]}{\mathrm{Tr}(\hat{a}\hat{\rho} \hat{a}^{\dagger})},
\end{equation}
where $\hat{\rho}=\hat{\rho}(t)$, $\hat{a}^{\dagger}$ is the associated creation operator, and $\tau=0$ corresponds to simultaneous time. Exchanging the creation/annihilation operators with phonon equivalents results in a corresponding phonon detection probability.

In Fig.~\ref{g1a_b} (a) $g^{(1)}_a$ tracks the monomer excited state for the four different system reorganization energies. In all cases Rabi oscillations persist through the entire simulation, and the period of the oscillation is dependent on the displacement, $\Delta$, and subsequent reduction of Franck-Condon overlap resulting in a hindered transition into the $\ket{e,0}$ state. Fig~\ref{g1a_b} (b) and (c) present the introduction of the bath through the bath reorganization energy, $\eta$. The addition of phonon dissipation results in the formation of a steady state in all cases apart from when the displacement of the excited state is zero. When $\lambda=0$ the ground state Boltzmann distribution is able to move perfectly to an equivalent Boltzmann distribution in the excited state resulting in a Rabi oscillation. When $\lambda>0$ and with a bath reorganization energy of $\eta$ a steady state forms in approximately $5\,\mathrm{ps}$, whereas for $2\eta$, it takes approximately $2.5\,\mathrm{ps}$.

Figure~\ref{g1a_b} (d), (e) and (f) are the equivalent simultaneous time detection probabilities for a phonon. It is clear that there is a definite minimum value of phonon transfer present within the monomer system which occurs when $\lambda=0$. Any increase in $\lambda$ (forcing the system away from equilibrium) results in increased VR and a linear increase in the $g^{(1)}_b$. Excluding the $\lambda=0$ case, all other values of system reorganization energy also exhibit the beating pattern of the electronic Rabi oscillation reflecting the departure from equilibrium. Just as in Fig.~\ref{g1a_b} (b), and (c), the results for (e) and (f) present a dramatic reduction in the formation time of the steady state when the bath is introduced. 

The non-normalised, second-order correlation function
\begin{equation}
    G_{ab}^{(2)}(t,\tau)=\mathrm{Tr}\left[\hat{a}^{\dagger}\hat{a}\exp(\mathcal{L}\tau)(\hat{b}\hat{\rho} \hat{b}^{\dagger})\right],\label{second_order_non_norm}
\end{equation}
where $\hat{\rho}=\hat{\rho}(t)$, represents the correlation of a photon followed by a phonon. 
%and $\tau=0$ corresponds to simultaneous time. 
The exchange of these operators results in a phonon-photon correlation $G^{(2)}_{ba}$, and versions containing purely $\hat{a}^{(\dagger)}$ and $\hat{b}^{(\dagger)}$ operators result in photon-photon and phonon-phonon correlations, $G^{(2)}_{aa}$ and $G^{(2)}_{bb}$, respectively.

\begin{figure*}[ht]
\centering
  \includegraphics[width=\textwidth]{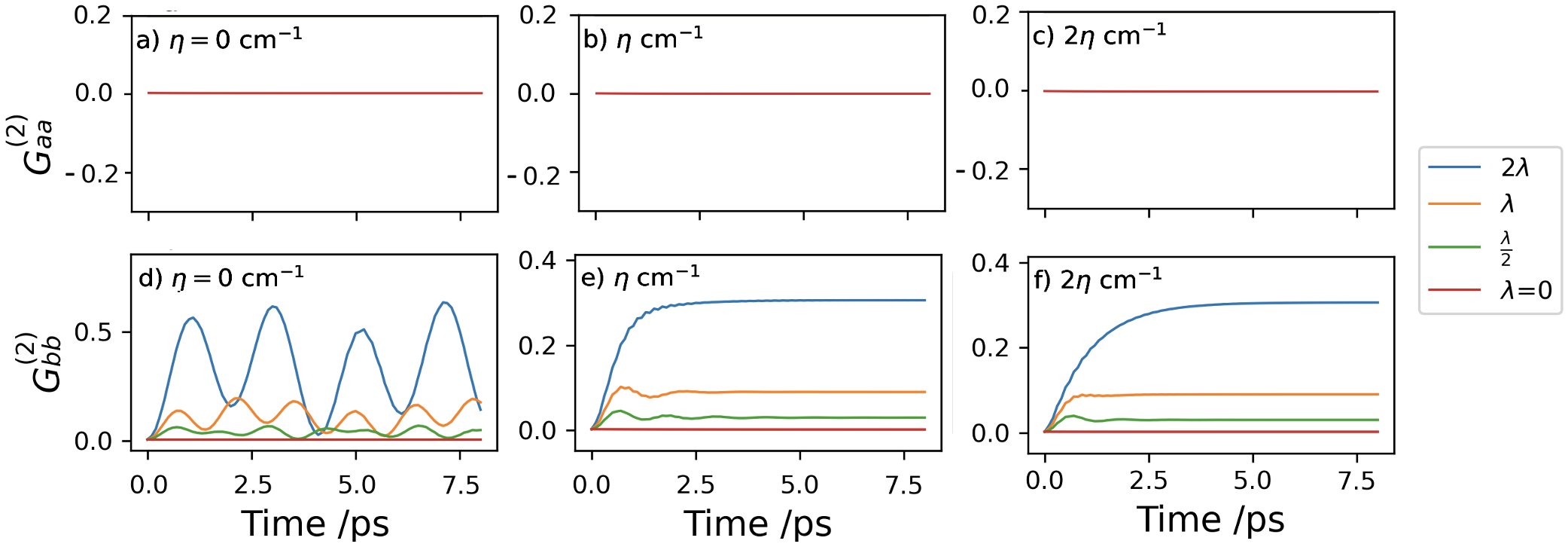}
  \caption{(a), (b), (c): $G^{(2)}_{aa}(t,\tau=0)$ second order non-normalised correlation function. (d), (e), (f): $G^{(2)}_{bb}(t,\tau=0)$ second order non-normalised correlation function. Both include scanning over bath and system reorganization energies $\eta$ and $\lambda$.}\label{g2aa_bb}
\end{figure*}

Figure \ref{g2aa_bb} (a)--(c) demonstrate the underlying physics of the chosen model. Based on the chosen monomer system, at simultaneous time $(t,\tau=0)$ it is impossible to have a double excitation as there is no electronic doubly excited state. As such, we must have a constant zero probability of two photon detection. In contrast, the system has many accessible vibrational states,  so simultaneous detection of two phonons is possible [Fig.~\ref{g2aa_bb} (d)--(f)].

\begin{figure*}
\centering
  \includegraphics[width=\textwidth]{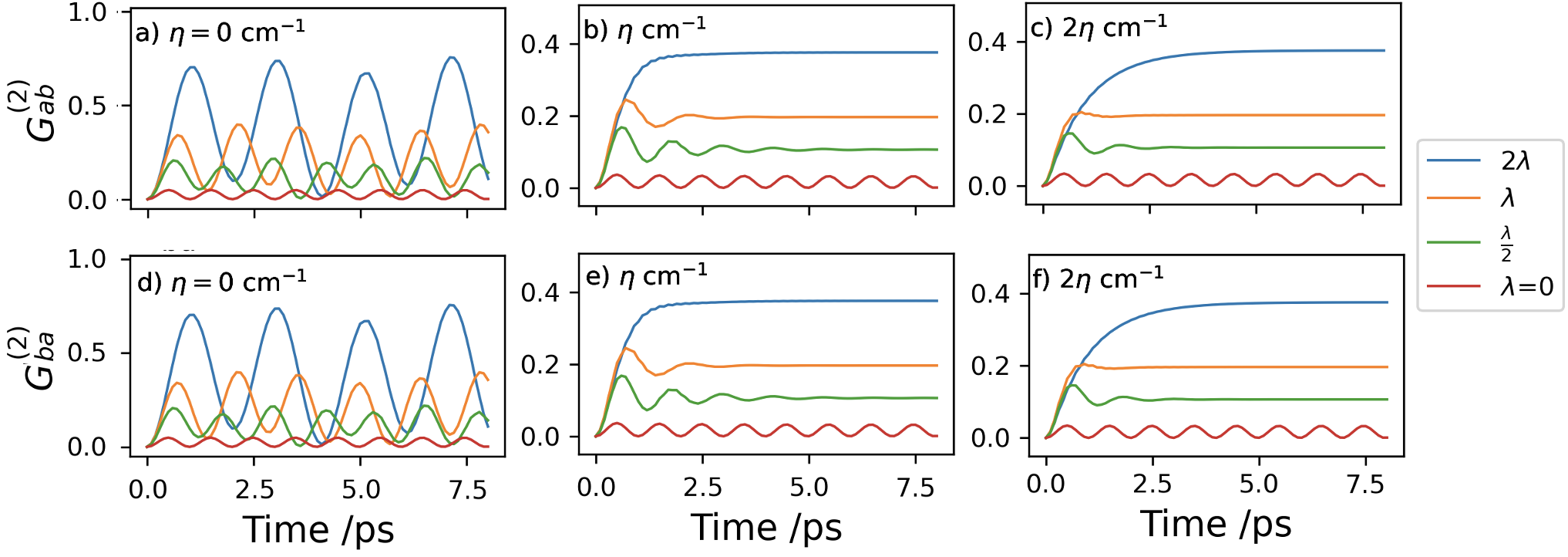}
  \caption{(a), (b), (c): $G^{(2)}_{ab}(t,\tau=0)$ non-normalised cross-correlation function at simultaneous time. (d), (e), (f): $G^{(2)}_{ba}(t,\tau=0)$ non-normalised cross-correlation function at simultaneous time. Both include scanning over bath and system reorganization energies $\eta$ and $\lambda$.}\label{g2ab_ba_unnorm}
\end{figure*}

Similarly, Fig.~\ref{g2ab_ba_unnorm} (a)--(e), are a demonstration of the nature of the second order cross-correlation functions. When at simultaneous time, the order of operations is irrelevant.  As such the simultaneous time, $\tau=0$, cross-correlations are equivalent and a superimposition of the photon detection Rabi oscillations on the behaviour of the phonon detection probabilities. 

\subsection{Normalization}

\begin{figure*}
\centering
  \includegraphics[width=\textwidth]{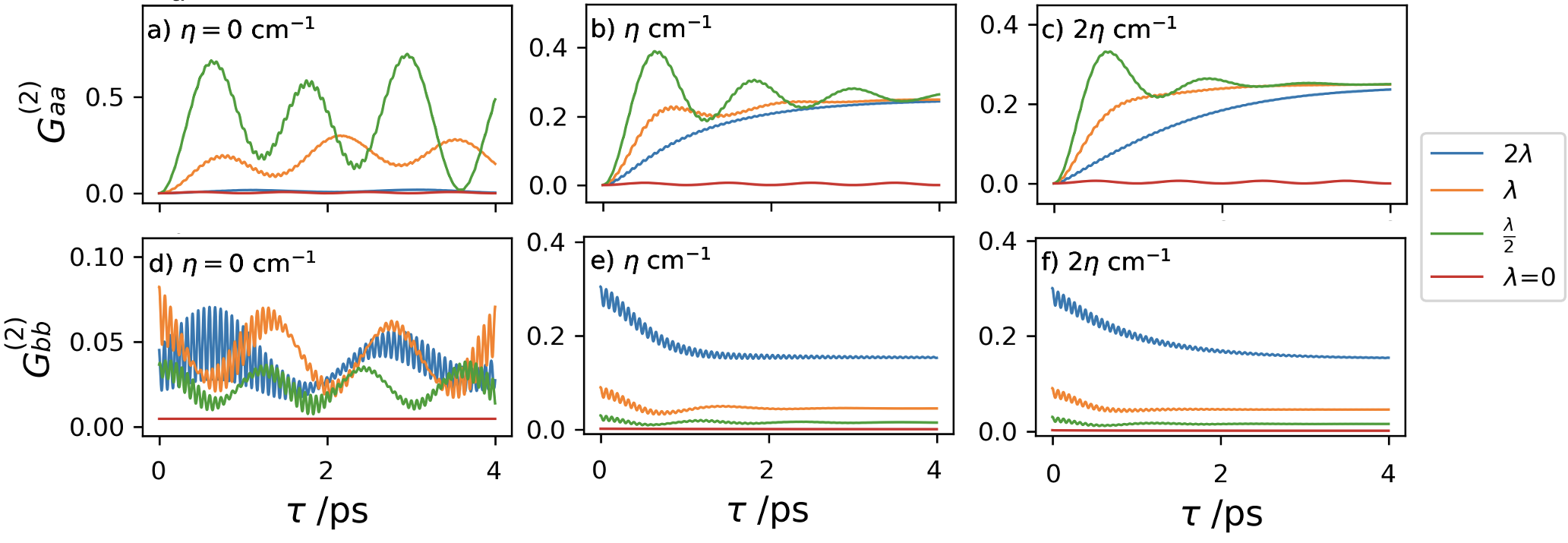}
  \caption{(a), (b), (c): $G^{(2)}_{aa}(t=3.5,\tau)$ non-normalised second order correlation function. (d), (e), (f): $G^{(2)}_{bb}(t=3.5,\tau)$ non-normalised second order correlation function. Both include scanning over the bath and system reorganization energies $\eta$ and $\lambda$, and both start the $\tau$ evolution from $t=3.5$ps.}\label{g2aa_bb_t_3.5}
\end{figure*}

\begin{figure*}
\centering
  \includegraphics[width=\textwidth]{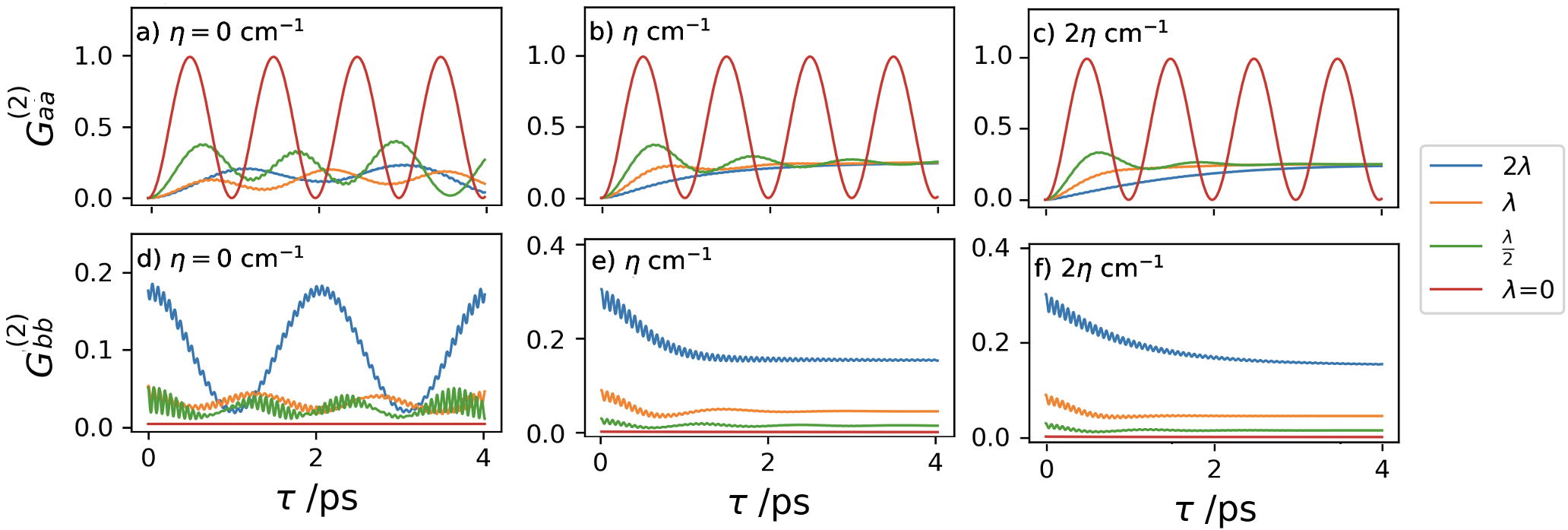}
  \caption{(a), (b), (c): $G^{(2)}_{ab}(t=4.0,\tau)$ non-normalised second order correlation function. (d), (e), (f): $G^{(2)}_{ba}(t=4.0,\tau)$ non-normalised second order correlation function. Both include scanning over the bath and system reorganization energies $\eta$ and $\lambda$, and both start the $\tau$ evolution from $t=4.0$ps.}\label{g2aa_bb_t_4.0}
\end{figure*}

\begin{figure*}
\centering
  \includegraphics[width=\textwidth]{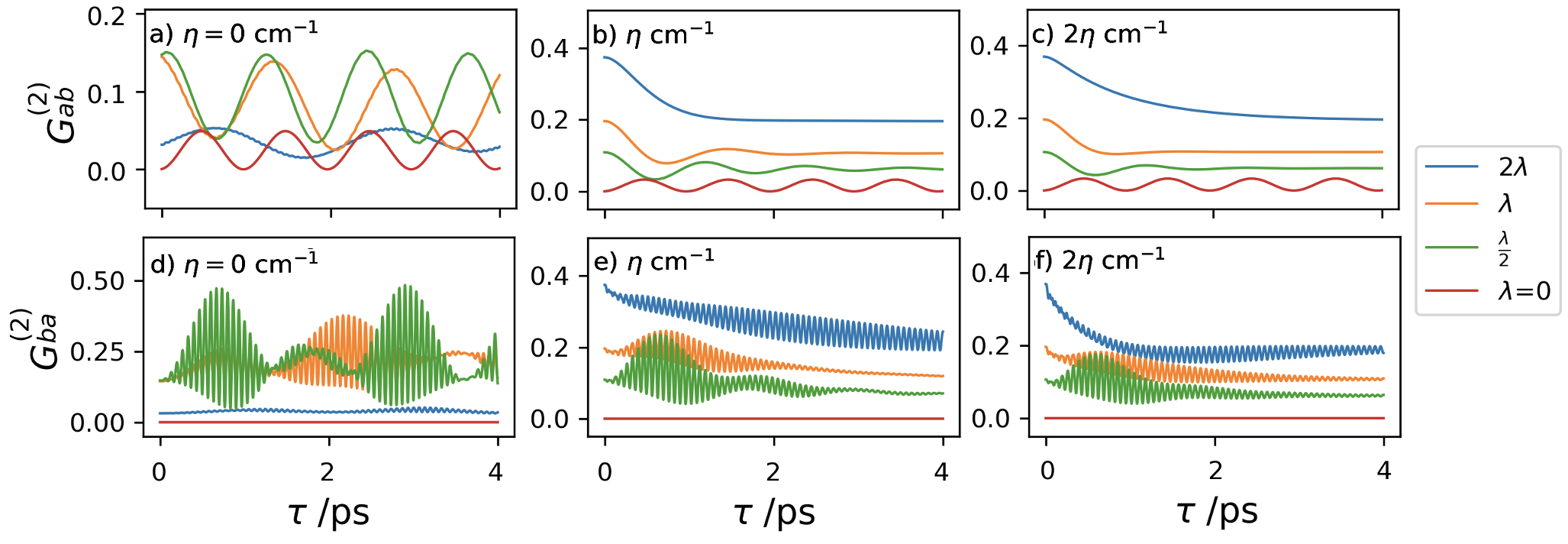}
  \caption{(a), (b), (c): $G^{(2)}_{ab}(t=3.5,\tau)$ non-normalised cross-correlation function. (d), (e), (f): $G^{(2)}_{ba}(t=3.5,\tau)$ non-normalised cross-correlation function. Both include scanning over the bath and system reorganization energies $\eta$ and $\lambda$, and both start the $\tau$ evolution from $t=3.5$ps.}\label{g2ab_ba_t_3.5}
\end{figure*}

\begin{figure*}
\centering
  \includegraphics[width=\textwidth]{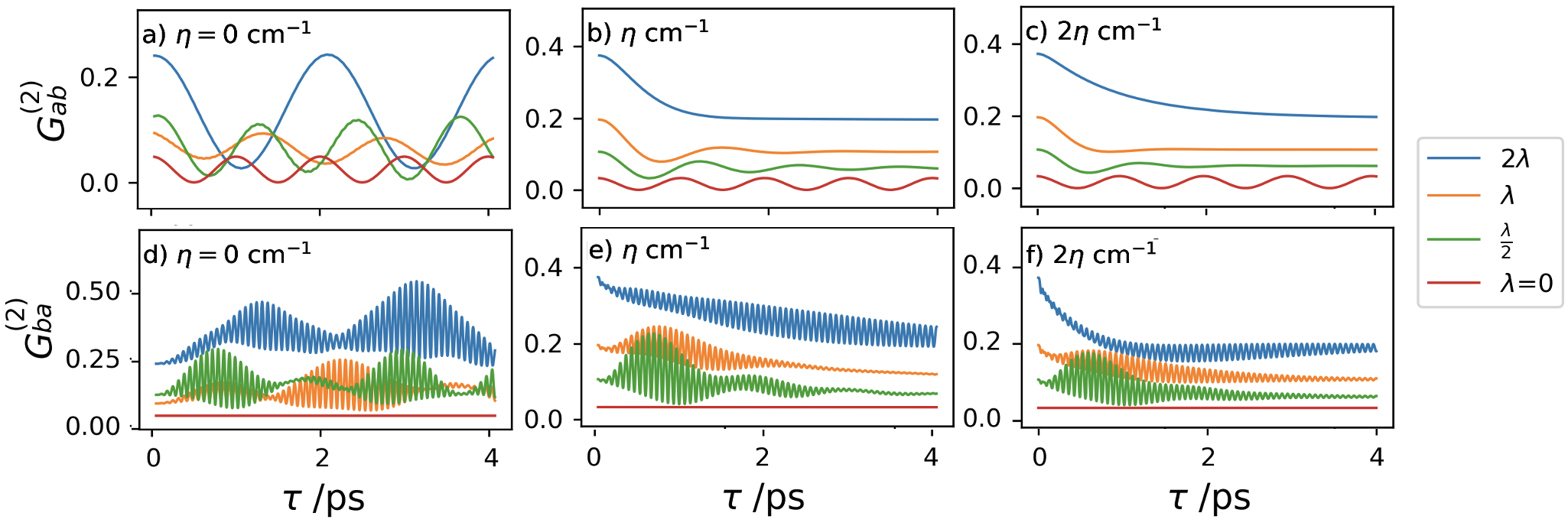}
  \caption{(a), (b), (c): $G^{(2)}_{ab}(t=4.0,\tau)$ non-normalised cross-correlation function. (d), (e), (f): $G^{(2)}_{ba}(t=4.0,\tau)$ non-normalised cross-correlation function. Both include scanning over the bath and system reorganization energies $\eta$ and $\lambda$, and both start the $\tau$ evolution from $t=4.0$ps.}\label{g2ab_ba_t_4.0}
\end{figure*}

Figures \ref{g2aa_bb_t_3.5}-\ref{g2ab_ba_t_4.0}, present non-normalised second-order correlation functions at two different values of $t$. The $0.5\,\mathrm{ps}$ increment corresponds to one half of the period for the non-displaced ($\lambda=0$) Rabi oscillation cycle: moving from $3.5$ to $4.0\,\mathrm{ps}$ goes from a trough in the photon detection probability to a peak. This is crucial for $\tau$ dependent results, as the starting point of the evolution dictates the starting amplitudes for all of the subsequent correlations, which can tend to favour or disfavour certain modes.

Within all of the results, there are two fundamental oscillatory modes: the major electronic Rabi oscillation, and the minor vibrational system mode frequency. The electronic Rabi oscillation is a result of the likelihood of photon emission from either the ground and excited states and is periodic due to the continuous driving by the laser field. The vibrational oscillation is a consequence of changes in the excited state wavepacket population with respect to the ground state Boltzmann distribution, corresponding to phonon transitions. 

\begin{figure*}[ht]
\centering
  \includegraphics[width=\textwidth]{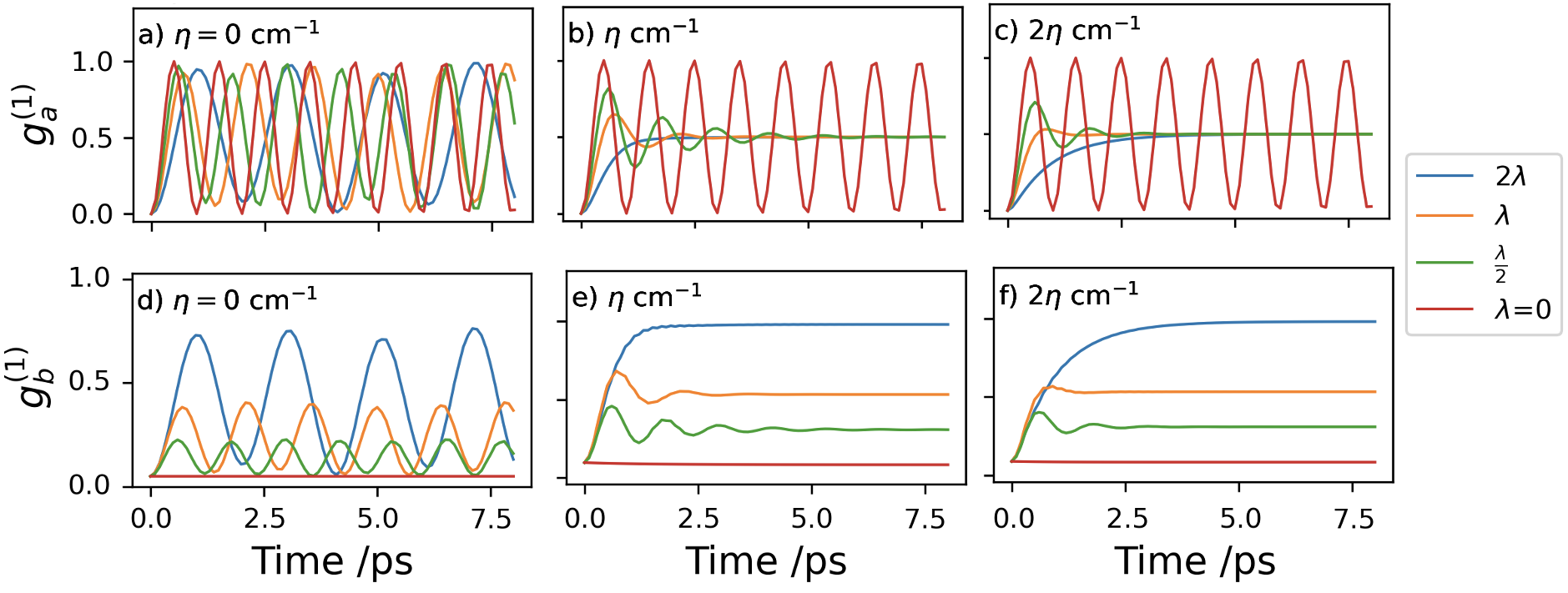}
  \caption{(a), (b), (c): $g^{(1)}_{a}(t,\tau=0)$ correlation function corresponding to photon detection probability. (d), (e), (f): $g^{(1)}_{b}(t,\tau=0)$ correlation function corresponding to phonon detection probability. Both include scanning over bath and system reorganization energies $\eta$ and $\lambda$.}\label{g1a_b}
\end{figure*}

Based on the excited state population profile within Fig.~\ref{g1a_b} (a), at a specific time each system is at a different position within its electronic Rabi oscillation and this determines the relative strength of further evolution. In figs \ref{g2aa_bb_t_3.5}, and  \ref{g2ab_ba_t_3.5}, $t=3.5$ps and $\frac{\lambda}{2}$, and $\lambda$ are the closest to their maximum value, whereas, $2\lambda$, and $\lambda=0$ are closest to their minimum. This minimises the impact of the electronic Rabi oscillation within the latter which consequently improves the resolution of the much smaller vibrational contributions. This is especially clear in the appearance of a minor oscillation at the system mode frequency in the $G^{(2)}_{ab}$ for $\eta=0$ with system reorganization of $2\lambda$ in Fig.~\ref{g2ab_ba_t_3.5} (a). The opposite is true (obscuring vibrational contributions) in figs \ref{g2aa_bb_t_4.0}, and \ref{g2ab_ba_t_4.0}, where $t=4.0\,\mathrm{ps}$. This motivates our choice to normalise either at the steady state value (for $\eta>0$), or at a time corresponding to an amplitude of one half, equivalent to the steady state average (for $\eta=0$), resulting in a unique time value for each system reorganization energy. 

\onecolumngrid

\vspace{2em}

\section{References}
%\bibliography{SI_bib}

\pagebreak